%% file: Main.tex
\documentclass[journal,onecolumn]{IEEEtran}

\usepackage[noadjust]{cite}
\usepackage{amsmath,amssymb,amsfonts,amsthm,mathtools}
\usepackage{graphicx}
\usepackage{textcomp}
\usepackage{url}

\usepackage{algpseudocode}
\usepackage[acronym,toc,shortcuts]{glossaries}
\usepackage{color}
\usepackage{soul}
\usepackage{framed}
\usepackage{array}
\usepackage{multirow}

\usepackage[utf8]{inputenc}   
\usepackage[T1]{fontenc} 

\usepackage[linesnumbered,ruled,vlined]{algorithm2e}
\usepackage[pdftex,bookmarks=true,colorlinks=true,linkcolor=blue,filecolor=blue, urlcolor=blue, citecolor=blue]{hyperref}
\usepackage{bookmark}

\usepackage{subcaption}

\makeglossaries
\input{glossaries}

\begin{document}

\title{On the Achievable Max-Min User Rates in Multi-Carrier Centralized NOMA-VLC Networks}

\author{Omar~Maraqa,  Umair~F.~Siddiqi,~\IEEEmembership{(Member,~IEEE)}, Saad Al-Ahmadi, and Sadiq~M.~Sait,~\IEEEmembership{(Senior~Member,~IEEE)}%

\thanks{O.~Maraqa and Saad~Al-Ahmadi are with the Department of Electrical Engineering, King Fahd University of Petroleum \& Minerals, Dhahran-31261, Saudi Arabia (e-mails: {g201307310,saadbd}@kfupm.edu.sa).}%
\thanks{Umair~F.~Siddiqi and Sadiq~M.~Sait are with the center of Communications and IT Research, Research Institute, King Fahd University of Petroleum \& Minerals, Dhahran-31261, Saudi Arabia (e-mail: {ufarooq,sadiq}@kfupm.edu.sa).}%
\thanks{Digital Object Identifier: https://doi.org/10.3390/s21113705}%
}

\markboth{May, 2021: Accepted by Sensors Journal, MDPI Publisher.}%
{Maraqa \MakeLowercase{\textit{et al.}}: On the Achievable Max-Min User Rates in Multi-Carrier Centralized NOMA-VLC Networks}

\maketitle

\begin{abstract}
Visible light communications (VLC) is gaining interest as one of the enablers of short-distance, high-data-rate applications, in future beyond 5G networks. Moreover, non-orthogonal multiple-access (NOMA)-enabled schemes have recently emerged as a promising multiple-access scheme for these networks that would allow realization of the target spectral efficiency and user fairness requirements. The integration of NOMA in the widely adopted orthogonal frequency-division multiplexing (OFDM)-based VLC networks would require an optimal resource allocation for the pair or the cluster of users sharing the same subcarrier(s). In this paper, the max-min rate of a multi-cell indoor centralized VLC network is maximized through optimizing user pairing, subcarrier allocation, and power allocation. The joint complex optimization problem is tackled using a low-complexity solution. At first, the user pairing is assumed to follow the divide-and-next-largest-difference user-pairing algorithm (D-NLUPA) that can ensure fairness among the different clusters. Then, subcarrier allocation and power allocation are solved iteratively through both the Simulated Annealing (SA) meta-heuristic algorithm and the bisection method. The obtained results quantify the achievable max-min user rates for the different relevant variants of NOMA-enabled schemes and shed new light on both the performance and design of multi-user multi-carrier NOMA-enabled centralized VLC networks.
\end{abstract} 


\section{Introduction} \label{sec:intro}

Wireless optical communications, particularly \ac{VLC}, has emerged as a bandwidth-abundant, secure, and cost-effective communications technology. It complements the existing \ac{RF} systems or even replaces them for some indoor and outdoor applications, such as conference and exhibition halls, office rooms, trains and airplane cabins, and so forth. It can also be deployed outdoors in \ac{V2V} and \ac{V2X} applications, as well as applications which are short-range, and which have a very high data rate in \ac{B5G} networks. \ac{VLC} is based on the principle of modulating light  from \ac{LEDs} without any adverse effects on the human eye and at the required illumination levels to transmit data. Clearly, this provides an excellent opportunity to utilize the existing illumination infrastructure for very high-speed and secure wireless communications~\cite{8528460}.

In a conventional distributed \ac{VLC} architecture, a set of $L$ transmit \ac{LEDs} communicate with a set of $N$ users in a way that each \ac{LED} acts as an \ac{AP} that serves its own subset of the $N$ users. Besides, all the access points (APs) contain a base-band unit followed by an optical front-end and connect to each other through the data backbone, as well as the electrical grid. While in \ac{C-LiAN} architecture, the base-band processing unit of each \ac{LED} is retracted and grouped in a central pool that is managed by a centralized controller, where the \ac{LEDs} are only responsible for the optical transmission/reception. In this way, the complexity and the associated cost of each \ac{VLC}-enabled \ac{LED} can be reduced. In addition, efficient joint-processing algorithms of signals from different \ac{LEDs} can be implemented to enhance resource management, scheduling, interference management, handover, and so forth in the network. The centralized controller is responsible for collecting the \ac{CSI} of the users, the traffic load status of the \ac{LEDs}, and the \ac{QoS} requirements of the users~\cite{8037964}.

In the past generations of wireless networks, the wisdom of \ac{OMA} schemes, such as \ac{TDMA}, \ac{FDMA}, \ac{CDMA}, \ac{OFDMA}, and \ac{WDMA} in both \ac{RF} and \ac{VLC} networks was dominant~\cite{8698841}. In these schemes, the different users are allocated orthogonal frequency channels or time slots or spreading codes, or wavelengths so that they do not interfere with one another as they access the network. However, the orthogonality requirement poses a limit on the number of users that may access the network resources, and hence reduces the overall spectral efficiency and increases the network latency. On the other hand, \ac{NOMA} schemes allow several users to share the same resource (e.g., a frequency/time resource block) and separate these users in other domains with additional receiver complexity. When the power domain is used to separate the users, it is referred to as the \ac{PD-NOMA}~\cite{islam2017power} scheme (in this paper, we refer to \ac{PD-NOMA}  and \ac{NOMA} interchangeably. Hence, wherever \ac{NOMA} is used after this point in the paper, it always refers to \ac{PD-NOMA}). The \ac{NOMA} concept was first introduced in~\cite{saito2013non} to enhance the spectral efficiency of wireless cellular networks, and it has its roots in broadcast channels in  multi-user information theory~\cite{liu2017non}, though with additional constraints on the users' target rates in addition to the typical transmit power constraint~\cite{maraqa_nomasurvey2020}.

What makes \ac{NOMA} interesting for adoption in indoor \ac{VLC} networks are the followings~\cite{8528460,7342274}: (i) As the current off-the-shelf \ac{LEDs} have limited bandwidth, this necessitates the adoption of spectrally efficient schemes, like the NOMA-enabled schemes, to attain the desired high data rates in \ac{VLC} networks, (ii) under typical illumination constraints, \ac{VLC} experiences relatively high \ac{SNR} conditions where it is well-known that  the \ac{NOMA} scheme outperforms the orthogonal counterparts in that particular region, and (iii) the quasi-static nature of the propagation channel in low mobility and indoor \ac{VLC} networks allows for more reliable estimation of the channel gains for subsequent \ac{NOMA} scheme power allocation in the \ac{LEDs}.

It is worth noting that applying \ac{NOMA} to a large number of users in a single carrier system is not that feasible due to the error propagation, the high complexity of the architecture, and the additional system overhead for coordinating the users' CSI feedback. Hence, multi-carrier \ac{NOMA}-enabled systems with efficient user-pairing techniques have emerged, where users are divided into clusters, and \ac{NOMA} is applied among each cluster. Each cluster can be served through one or more subcarriers. In this paper, user pairing, subcarrier allocation, and power allocation in \ac{DCO-OFDM} indoor \ac{NOMA}-enabled centralized \ac{VLC} networks are investigated.  Next, we provide a discussion for the most-utilized methods that can solve user pairing, subcarrier allocation, and power allocation problems, and we justify the use of the methods that we adopted in our proposed heuristic-based solution.

In the \ac{NOMA} literature, there are different proposed user-pairing algorithms, such as (i)~the random user-pairing algorithm~\cite{8352621}, (ii)~cognitive radio (CR)-inspired-\ac{NOMA} pairing algorithm~\cite{7273963}, (iii)~\ac{NLUPA}~\cite{7273963}, (iv)~divide-and-NLUPA (D-NLUPA)~\cite{8352621}, (v)~user-pairing algorithms that utilize the matching theory, such as the Hungarian method and Gale-Shapley algorithm~\cite{8278151}, and (vi)~the exhaustive search user-pairing algorithm~\cite{8278151}. Among these algorithms, the D-NLUPA algorithm is adopted in this work, as it is attractive for the max-min user rate requirement~\cite{8352621}. The reason is that by re-distributing the rate gains between the clusters and achieving cluster fairness, the rate gap between the users' rates will reduce and this reduces the search space, and thus the algorithm complexity, for reaching the max-min user rate point for all the users in the network.

The subcarrier allocation problem is typically a \ac{NP}-hard optimization problem in both the number of users and the number of channels to be allocated~\cite{6514044}. The optimal resource allocation problem is known to be an \ac{NP}-hard problem even for the single carrier case~\cite{6514044}. The NP class contains problems that have the characteristic that only non-deterministic algorithms can solve them in polynomial-time. However, we can verify a given solution to these problems in polynomial time. The non-deterministic algorithms solve a problem by evaluating possible guesses of the solution. A problem is NP-hard if it is at least as hard as the hardest problem in NP, and all NP problems are reducible to it. In NP-hard problems, the search space size is so large that an exhaustive search is infeasible. Therefore, we determine approximate solutions to NP-hard problems. Evolutionary algorithms (EAs) are a popular approach to solve NP-hard problems~\cite{Sait:1999}.  Simulated Annealing (SA) is a well-known EA that has demonstrated excellent performance in solving optimization problems in various applications, including wireless communications~\cite{TIAN1995629, doi:10.1002/ett.3850, 8715449, Sait:1999}. In this work, we use the SA algorithm to allocate the subcarriers to user-pairs. For the power allocation, the bisection method is commonly used to optimally solve the power allocation problem for max-min user rate optimization in the \ac{NOMA} literature~\cite{7069272,6991587}.

\subsection*{Related Work}
When it comes to rate-optimal \ac{NOMA}-enabled \ac{VLC} networks, the majority of works have considered maximizing the system sum-rate metric~\cite{yang2017fair,tahira2019optimization,8742679,8936365,9075277,fu2018enhancedo,8669970,8554281}, and this is detailed in Table~\textrm{VIII} of the survey by Maraqa~et~al. \cite{maraqa_nomasurvey2020}. In particular, early works on rate-optimal \ac{NOMA}-enabled \ac{VLC} networks have considered a single-LED, single-carrier setup as \mbox{in~\cite{yang2017fair, tahira2019optimization, 8742679, 8936365, 9075277}}. Then, the research evolved to consider a single-LED multi-carrier setup as in~\cite{fu2018enhancedo, 8669970}. Later, a few works have considered a multi-LED, multi-carrier setup as \mbox{in~\cite{8554281, 7792590, 9062301}}. Among those, only two works have considered the optimization of another common fairness measure, which is the max-min achievable user rate alongside maximizing the system sum-rate metric~\cite{7792590,9062301}. This is referred to as the reason that the optimization of the max-min user rate tends to reduce the total throughput of the users, as compared to sum-rate maximization; however, it guarantees fairness among the users. This is an appealing feature for the operation of \ac{umMTC} and \ac{IoE} in \ac{B5G} systems that need reliable connectivity with generally low data-rate requirements~\cite{8766143}. The work in~\cite{7792590} has considered the maximization of the minimum rate for \ac{NOMA}-enabled \ac{VLC} networks, where an approximation of the objective function was utilized to convexify the optimization problem, and a \ac{GP} algorithm is used to adjust the power allocation coefficients. The maximization of the minimum user rate for the cell-edge users in \ac{NOMA}-enabled \ac{VLC} networks in~\cite{9062301}, using hybrid \ac{NOMA} and/or linear \ac{ZF} pre-coding schemes, was approached by numerical convex optimization of the standard determinant maximization. However, both aforementioned works have considered the optimization of both the power allocation and user grouping. Neither of them has analyzed the effect of subcarrier allocation on the max-min user rate performance of \ac{NOMA}-enabled \ac{VLC} networks, which is the main objective of this work.

\begin{table*}[!b]
\centering
\caption{List of used variables.}
\label{Table: List of used variables}
\begin{tabular}{|c|l|}
\hline
\textbf{Variable Name} & \textbf{Variable Description}\\ \hline

\multicolumn{2}{|c|}{\textbf{System Model's Variables}}  \\\hline

$\mathcal{L}$= $\{l_0,l_1, ...,l_{L-1}\}$ & A set to describe the total number of \ac{LEDs} \\\hline

$L$ & The total number of \ac{LEDs} \\\hline

$\mathcal{K}$= $\{s_0,s_1,.., s_{K-1}\}$& A set to describe the total number of subcarriers\\\hline

$K$& The total number of subcarriers\\\hline

$\mathcal{N}$= $\{u_0,u_1, ...,u_{N-1}\}$& A set to describe the total number of users\\\hline

$N$& The total number of users\\\hline

$N_0$& The total number of users served by the \ac{LED} $l_0$ \\\hline

$\varphi_{l_0,u_s}$& The angle of irradiance between the \ac{LED} $l_0$ and the user $u_s$ \\\hline

$\psi_{l_0,u_s}$& The angle of incidence between the \ac{LED} $l_0$ and the user $u_s$\\\hline

$u_s$ and $u_w$& A strong user and a weak user of the \ac{LED} $l_0$\\\hline

$d_{l_0,u_s}$& The distance between the \ac{LED} $l_0$ and the user $u_s$\\\hline

$\Psi_{1/2}$ & The \ac{FoV} semi-angle of the user $u_s$ \\\hline

$m$ & The order of Lambertian emission\\\hline

$\phi_{1/2}$ & The semi-angle of the \ac{LED} $l_0$\\\hline

$A_p$ & The area of the \ac{PD} for the user $u_s$\\\hline

$T_s(\psi_{l_0,u_s})$ & The optical filter gain\\\hline

$\chi$  & The refractive index\\\hline

$h_{l_0,u_s}$& The channel gain between the \ac{LED} $l_0$ and the user $u_s$ \\\hline

$a_s$& The power allocation coefficient for the strong user \\\hline

$a_w$& The power allocation coefficient for the weak user \\\hline

$P_e^k$ & The electrical signal power per subcarrier of a \ac{LED} \\\hline

$P_e$ & The electrical signal power of a \ac{LED} \\\hline

$P_o$ & The optical transmit power at the output of a \ac{LED} \\\hline

$\iota= P_o/\sqrt{P_e}$ &The ratio between the electrical signal power and the optical transmit power \\\hline

$\kappa$ & The optical to electrical conversion efficiency of the \ac{PDs} \\\hline

$\sigma_k^2=Z_o B_L/K$ &  \begin{tabular}[l]{@{}l@{}} The power of equivalent AWGN where $Z_o$ denotes the noise power spectral density and \\$B_L$ denotes the base-band modulation bandwidth \end{tabular}\\\hline

$\gamma_s^k$ and $R_s$ & The SINR and the achievable rate of the strong user \\\hline

$\gamma_w^k$ and $R_w$ & The SINR and the achievable rate of the weak user \\\hline

$R_j$ & \begin{tabular}[l]{@{}l@{}} The achievable rate of an arbitrary user ($j$-th user) of an arbitrary \ac{LED} in the network\\ (i.e., $R_j$ can be a strong user or a weak user in arbitrary user pair)\end{tabular} \\\hline

$S_{l_i,k} \in \big\{0,1\big\}$ & A binary variable to  denote that a user is served by a \ac{LED} $l_i$ and a subcarrier $k$ \\\hline

$S_k \in \big\{0,1\big\}$ &  A binary variable to denote that a user is served by a subcarrier $k$ \\\hline

$S_{l_i}^j \in \{0,1\}$ & A binary variable to denote the user $u_j$  is served by \ac{LED} $l_i$ \\\hline

$\boldsymbol{\Gamma}(l_i)$ &  A set of all user-pair combinations for an arbitrary \ac{LED} $l_i$\\\hline

$K_{l_i}$ &  The maximum number of subcarriers for an arbitrary \ac{LED} $l_i$ \\\hline

\multicolumn{2}{|c|}{\textbf{Heuristic-based Solution's Variables}}  \\\hline 

$\lambda(u_j)$ & The most suitable \ac{LED} for a user $u_j \in \mathcal{N}$\\\hline

$f_1(l_i)$ &  A function that returns the number of users assigned to the \ac{LED} $l_i$\\\hline

$f_2(u_j)$ & \begin{tabular}[l]{@{}l@{}} A function that denotes the distance of the user $u_j$ from the \ac{LED} to which it \\is currently allocated  \end{tabular}\\\hline

$f_3(u_j)$ & A function that denotes the maximum distance of $u_j$ from any \ac{LED} \\\hline

$\mathcal{N}_{l_i}$, $N_{l_i}$ &  A vector that store the users of \ac{LED} $l_i$ after binding, the number of users in $\mathcal{N}_{l_i}$ \\\hline

$\mathbb{P}_j$ & A pair of users in $\boldsymbol{\Gamma}(l_i)$\\\hline

$X$ & A decision matrix that represents the solution of subcarrier allocation\\\hline

$\mathcal{C} \in [0,1]$ & A predefined constant \\\hline

$\mathcal{P}_1$ and $\mathcal{P}_2$ & The penalty factors of the penalty method\\\hline

\multicolumn{2}{|c|}{\textbf{Simulated Annealing Algorithm's Variables}}  \\\hline
$T_0$ and $T$ & The initial temperature and the current temperature\\ \hline
$\alpha$ & The rate of decrease in the temperature \\ \hline
$M_0$ and  $M$ & The initial and current value of the number of iterations in the Metropolis function \\ \hline
$\beta$ & The rate of increase in the number of iterations of the Metropolis function \\ \hline
$X$,$X_{\textnormal{current}}$, and $X_{\textnormal{best}}$ & The input solution, the current solution, and the best solution\\ \hline
$c_{\textnormal{current}}$, $c_{\textnormal{best}}$, and $c_{\textnormal{new}}$ & \begin{tabular}[l]{@{}l@{}}The costs (objection function value) of the current solution, the best solution,\\ and the new solution created in the Metropolis function \end{tabular}\\ \hline

\end{tabular}%
\end{table*}

Our contributions can thus be summarized as follows:
\begin{itemize}
\item The optimization of the achievable max-min user rates for \ac{NOMA}-enabled centralized \ac{VLC} is investigated through formulating a joint problem for the user pairing, the subcarrier allocation, and the power allocation. Then, a low complexity solution is~proposed.

\item The development of \ac{SA}-assisted algorithm for tackling the subcarrier allocation in the max-min user rate optimization problem. The obtained results are further verified using the \ac{TS} algorithm.

\item The implementation of both of the \ac{NOMA}-imposed schemes, where all the users are grouped into pairs, and the \ac{NOMA}-not-imposed schemes; and besides these, the investigation of the effect of the different network parameters on the achievable max-min user rate.
\end{itemize}

The remainder of the paper is organized as follows. In Section~\ref{Section:System Model}, we provide the system and channel models for the utilized \ac{C-LiAN} architecture. The formulation of the proposed max-min user rate optimization problem is analyzed in Section~\ref{Section: The Max-Min User Rate Optimization Problem}. In \mbox{Section~\ref{Section:Heuristic Algorithm}}, the details of the proposed heuristic-based solution for the formulated max-min user rate optimization are explained. Finally, simulation results, paper conclusions, and {future research directions} are given in  Section~\ref{Section:Results and Discussions}, Section~\ref{Section:Conclusions} and {Section~\ref{Section: Extensions and Future Work}}, respectively. Table~\ref{Table: List of used variables} summarizes the notations used in this paper.

\section{System and Channel Models}
\label{Section:System Model}

\begin{figure}[!ht]
\centering
\includegraphics[width=0.75\textwidth]{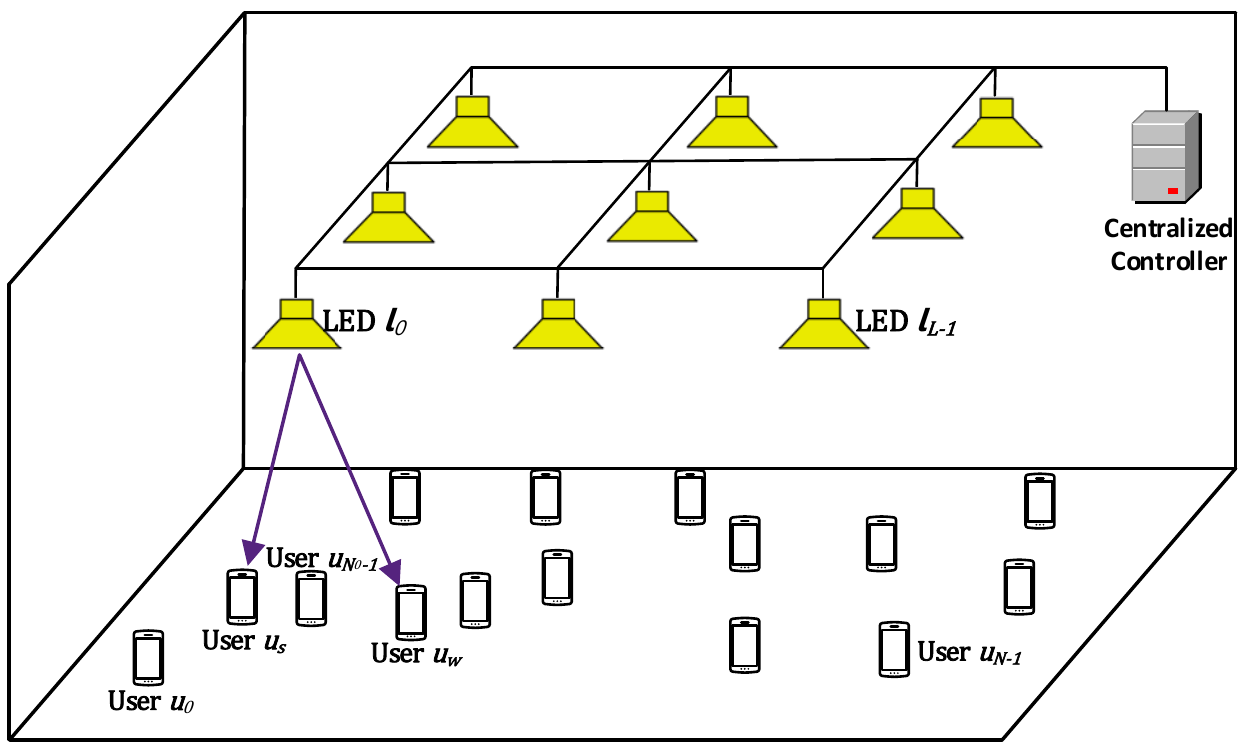}
\caption{The \ac{C-LiAN} architecture~\cite{8037964}.}
\label{fig:System model}
\end{figure}

A multi-user \ac{NOMA}-enabled \ac{C-LiAN} architecture~\cite{8037964} is shown in Fig.~\ref{fig:System model}. In this paper, we assume that the ceiling of the room contains uniformly distributed \ac{LEDs} denoted by $\mathcal{L}$ = $\{l_0, l_1, ..., l_{L-1}\}$, where $L$ is the number of \ac{LEDs}. Each \ac{LED} is assigned up to $K$ subcarriers that can be shared by several users, where the subcarriers of each \ac{LED} are denoted as $\mathcal{K}$ = $\{s_0, s_1,.., s_{K-1}\}$. In addition, the users are randomly located in the room, and denoted by $\mathcal{U}$ = $\{u_0,u_1, ..., u_{N-1}\}$; where the total number of users in the network is $N$. In multi-user systems, the transmission resources described in time, wavelength, and/or space are divided into resource units. A key challenge is the allocation of the resource units in a way that key performance metrics, such as user fairness and spectral efficiency, are fulfilled. This motivates allowing the served users or a subset of them to share the subcarriers or a subset of them. Hence, in each signaling interval, we assume that each pair of users can occupy one or more subcarrier(s) in only one \ac{LED}.

\begin{figure}[!ht]
\centering
\includegraphics[width=0.38\textwidth]{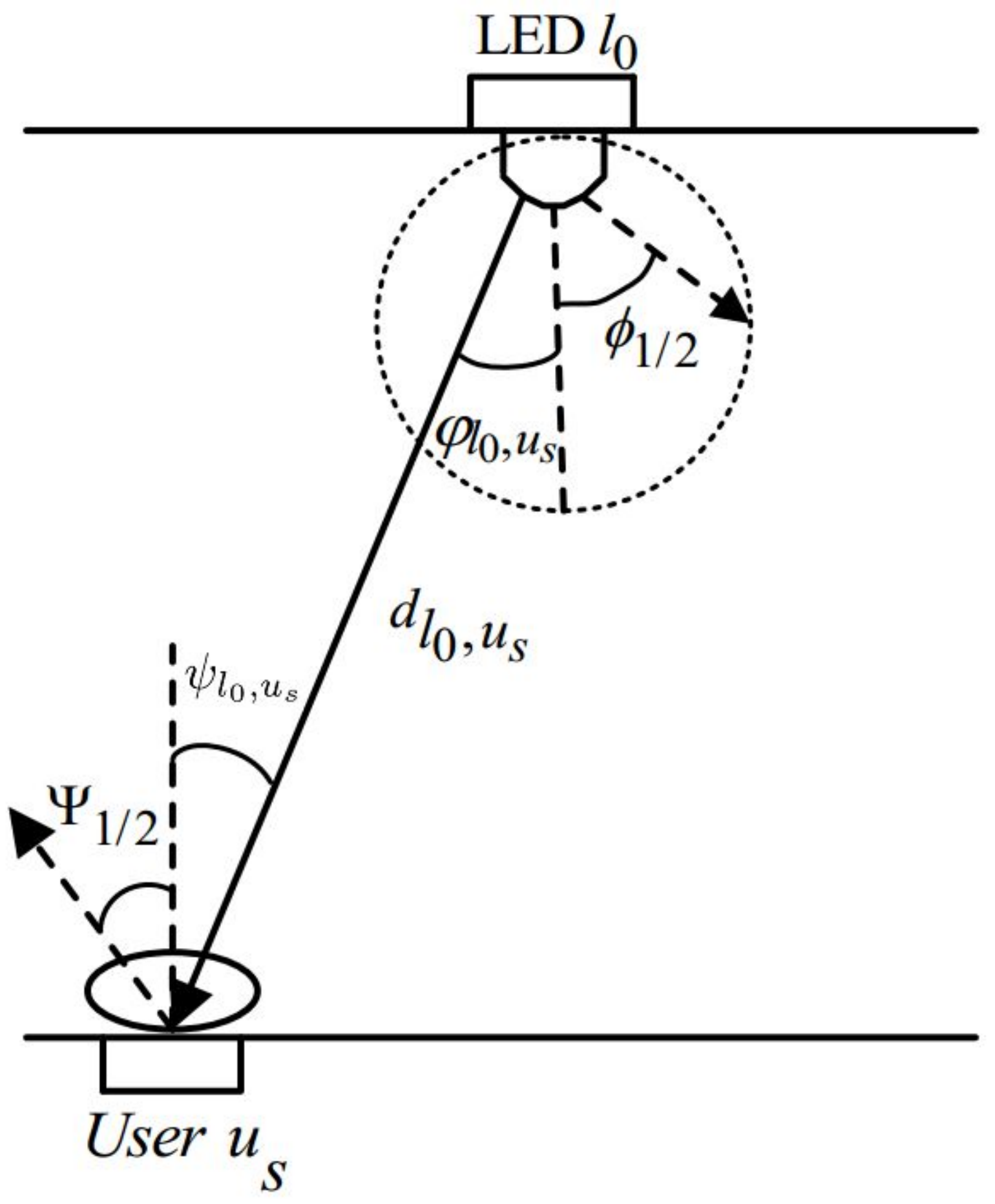}
\caption{The LoS channel model between LED $l_0$ and {user} $u_s$~\cite{8428395}.\label{fig:LoS channel model}}
\end{figure}

In this work, the \ac{LoS} optical channel model is adopted, since in a typical indoor environment, the strongest diffuse component is usually much weaker than the LoS component~\cite{5342325}. Hence, the channel gain between the \ac{LED} $l_0$ and user $u_s$ is given by~\cite{8715449}:

\begin{equation} \label{eq: 1}
h_{l_0,u_s}= \begin{cases} \frac{(m+1)A_p \chi^2 T_s(\psi_{l_0,u_s})}{2\pi d_{l_0,u_s}^2\sin^2(\Psi_{1/2})}\cos^m(\varphi_{l_0,u_s})\cos(\psi_{l_0,u_s}) \ \text{if }\\
\ \ \qquad \qquad 0 \leq \psi_{l_0,u_s} \leq \Psi_{1/2} \\
0, \qquad \qquad \psi_{l_0,u_s} \geq \Psi_{1/2} \end{cases}
\end{equation}

\noindent where $\varphi_{l_0,u_s}$ and $\psi_{l_0,u_s}$ denote, respectively, the angle of irradiance and angle of incidence between the \ac{LED} $l_0$ and user $u_s$. As shown in Fig.~\ref{fig:LoS channel model}, $d_{l_0,u_s}$ is the distance between the \ac{LED} $l_0$ and user $u_s$. $\Psi_{1/2}$ is the \ac{FoV} semi-angle of the user $u_s$. $m$ is the order of Lambertian emission and is equal to $\frac{-1}{\textrm {log}_2(\textrm {cos}(\phi_{1/2}))}$, where $\phi_{1/2}$ is the semi-angle of the \ac{LED} $l_0$. $A_p$ is the area of the \ac{PD} for user $u_s$, $T_s(\psi_{l_0,u_s})$ is the optical filter gain, and $\chi$ denotes the refractive~index.

We assume that served users by \ac{LED} $l_0$ are ordered according to their channel gains in ascending order as $|h_{l_0,u_0}|  \leq  |h_{l_0,u_1}|  \leq  ...\leq |h_{l_0,u_{N_0-1}}|$; where we assume here that the total number of served users by \ac{LED} $l_0$ is $N_0$. In general, if the user $u_0$ (i.e.,~the user that has the weakest instantaneous channel gain) of \ac{LED} $l_0$ occupies an arbitrary subcarrier (for example, the $k$-th subcarrier), then the \ac{SINR} of the user $u_0$ can be expressed as:

\begin{equation} \label{eq:2}
\gamma_{{u_0},k}=\frac{|h_{l_0,u_0}|^2 \kappa^2 a_0 P_{o,k}^2}{I_1 + I_2 +\iota^2 \sigma_k^2},
\end{equation}

\noindent where $a_0$ represents the power allocation factors for the user $u_0$ that is attached to \ac{LED} $l_0$. $P_{o,k}$ is the \ac{LED} optical transmit power per subcarrier and is equal to $P_o/(K-2)$ according to the \ac{DCO-OFDM} principle, where $P_o$ denotes the optical transmit power at the output of the \ac{LED}. The ratio between the electrical signal power and the optical transmit power can be expressed as $\iota= \frac{P_{o}}{\sqrt{P_{e}}}$, where $P_e$ denotes the electrical signal power. $\kappa$ denotes the optical-to-electrical conversion efficiency of the \ac{PDs}, {$\sigma_k^2$ is the power of the equivalent additive Gaussian noise} ({{{practical} indoor VLC networks are affected by both thermal noise due to the receiver pre-amplifier, and ambient light shot noise due to the possible sun-light through the windows and/or non-VLC indoor light sources. The total noise can be modeled as signal-independent Gaussian noise~\cite{7147818,ghassemlooy2017visible}), whose variance is the sum of the variances of these two noise components and equals to $Z_o B_L/K$; where $Z_o$ denotes the noise power spectral density. In the simulation section of this work, the value of $Z_o$ accounts for both thermal noise and light shot noise.}} $B_L$ denotes the base-band modulation bandwidth. In Equation~\eqref{eq:2}, the first term in the denominator, $I_1= \sum_{i=1}^{L-1} S_{l_i,k} |h_{l_i,u_0}|^2 \kappa^2 P_{o,k}^2$, represents the interference induced from the users that utilize the $k$-th subcarrier in the interfering \ac{LEDs} (i.e.,~inter-\ac{LED} interference) where $S_{l_i,k} \in \big\{0,1\big\}$ is a binary variable that denotes the user has received an interfering signal from a \ac{LED} $l_i$ on a subcarrier $k$. Likewise, the second term in the denominator, $I_2= \sum_{j=u_1}^{N_{0}-1} |h_{l_0,j}|^2 \kappa^2 a_j P_{o,k}^2$, represents the interference that remains after \ac{SIC} decoding (i.e.,~inter-user interference) where $a_j$ represents the power allocation factors for the rest of the users that is attached to \ac{LED} $l_0$. It is worthy to note that we assume in the above equation that multiple users can be served by one subcarrier of \ac{LED} $l_0$ (i.e.,~multiple users per cluster). For the rest of the paper, we consider the case where two users are served by the same subcarrier(s) (i.e.,~two users per cluster). Studying the system performance while considering multiple users per cluster is left for future research.

For the two-user pairing case, we use the \ac{D-NLUPA} scheme for pairing the users, as mentioned in \mbox{Section~\ref{sec:intro}}, where the strong user (i.e.,~user $u_s$) is close to the \ac{LED} with strong channel gain and the weak user (i.e.,~user $u_w$) is farther away from the \ac{LED} with weak channel gain so that the strong user can successfully decode and subtract the weak user signal before decoding its own signal (i.e.,~performing \ac{SIC}). In Fig.~\ref{fig:System model}, for \ac{LED} $l_0$, we can express the \ac{SINR} of the strong user, the \ac{SINR} of the weak, and the achievable rates of the strong and the weak users using a well-known lower bound for the capacity, given in~\cite{6636053,8428395}, as

\begin{gather}
\label{eq:5}
R_s=\frac{B_L}{K}  \sum_{k=0}^{K-1} S_k \log_2 \big(1+ \frac{e}{2\pi} \frac{ |h_{l_0,u_s}|^2 \kappa^2 a_s P_{o,k}^2 }{\sum_{i=1}^{L-1} S_{l_i,k} |h_{l_i,u_s}|^2 \kappa^2  P_{o,k}^2 +\iota^2 \sigma_k^2} \big),\\
\label{eq:6}
R_w=\frac{B_L}{K} \sum_{k=0}^{K-1} S_k \log_2  \big(1+ \frac{e}{2\pi} \frac{ |h_{l_0,u_w}|^2 \kappa^2 a_w P_{o,k}^2}{ \sum_{i=1}^{L-1} S_{l_i,k} |H_{l_i,u_w}|^2 \kappa^2  P_{o,k}^2 + |h_{l_0,u_s}|^2 \kappa^2 a_s P_{o,k}^2 +\iota^2 \sigma_k^2} \big),
\end{gather}

\noindent where $S_k \in \big\{0,1\big\}$ is a binary variable to denote that the user is served by a subcarrier $k$, and it is assumed that the two-user pair (user $u_s$ and user $u_w$) occupies one or more subcarrier(s) of the \ac{LED} $l_0$ and the power allocation factors satisfy the condition $a_s + a_w = 1$. Noting that the analysis in Equations~\eqref{eq:5}~and~\eqref{eq:6} represents the rate equations for one user-pair of \ac{LED} $l_0$, the rate equations for the rest of the user-pairs of \ac{LED} $l_0$ and for the user-pairs of any other \ac{LED} in the network can be analyzed similarly.

\section{The Max-Min User Rate Optimization Problem}
\label{Section: The Max-Min User Rate Optimization Problem}

In this section, we formulate a joint optimization problem for user pairing, subcarrier allocation, and power allocation that maximizes the minimum achievable rate of the served users. Let $R_j$ denote the achievable rate of an arbitrary user (i.e, $j$-th user) of an arbitrary \ac{LED} (i.e, \ac{LED} $l_i$) in the network (i.e.,~$R_j$ can be a strong user or a weak user in a user pair, as described in \mbox{Equations~(\ref{eq:5}) and (\ref{eq:6})}). Then, the max-min optimization problem can be expressed as: 
\begin{alignat}{3}
&\underset{\boldsymbol{\Gamma}(l_i), \mathcal{K}, a_s, a_w}{\text{maximin}} &\qquad&R_j,\label{eq:objective-2}\\
&\text{subject to} &  &   \sum_{k=0}^{K-1} S_k^j \geq 1, \ \forall j, \  0 \leq j \leq N_{l_i} -1 \label{eq:constraint-b-2}\\
&  &  & 0 \leq K_{l_i} \leq \frac{K}{2}-1, \label{eq:constraint-c-2} \\
&  &  & \sum_{i=0}^{L-1} S_{l_i}^j = 1, \ \forall j, \  0 \leq j \leq N_{l_i} -1 \label{eq:constraint-d-2} \\
&  &  &a_s + a_w \leq 1, \ \forall \text{pair} \in \boldsymbol{\Gamma}(l_i) \label{eq:constraint-e-2}
\end{alignat}

\noindent where $\boldsymbol{\Gamma}(l_i)$ is a set of all user-pair combinations of an arbitrary \ac{LED} $l_i$. $S_k^j \in \{0,1\}$ and $S_{l_i}^j \in \{0,1\}$ are binary variables to denote that the user $u_j$ is served by a subcarrier $k$ and the user $u_j$ is served by \ac{LED} $l_i$, respectively. By Equation~(\ref{eq:constraint-b-2}), we ensure that any user attached to \ac{LED} $l_i$ is served by at least one subcarrier, noting that the weak user and the strong user in a user-pair are served by the same number of subcarriers according to the \ac{NOMA} principle. By Equation~(\ref{eq:constraint-c-2}), the maximum number of subcarriers per \ac{LED} is limited to $\frac{K}{2}-1$ to be aligned with the \ac{DCO-OFDM} principle. By Equation~(\ref{eq:constraint-d-2}), we ensure that each user attached to \ac{LED} $l_i$ is served only by \ac{LED} $l_i$. By Equation~(\ref{eq:constraint-e-2}), we limit the values of the power allocation coefficients in a user-pair. The above optimization problem is of a combinatorial nature, for the user pairing and the subcarrier allocation, and has a non-linear objective function, and hence a non-convex \ac{NP}-hard optimization problem~\cite{4453890,6514044}. The optimal values of the power allocation coefficients can be determined using the bisection method~\cite{7069272}.

\section{The Heuristic-Based Solution for the Max-Min User Rate Optimization Problem}
\label{Section:Heuristic Algorithm}

The optimization problem in Equations~\eqref{eq:objective-2}--\eqref{eq:constraint-e-2} is an involved optimization problem, as stated before. Subsequently, the search space to handle the problem jointly is too large. Noting that performing an exhaustive search to solve this problem jointly is infeasible from a practical point of view, to reduce the complexity of this optimization problem, the user pairing is assumed to follow the \ac{D-NLUPA} algorithm, as mentioned before, where that pairing algorithm is attractive for the max-min user rate requirement~\cite{8352621}. Hence, the optimization problem in Equations~\eqref{eq:objective-2}--\eqref{eq:constraint-e-2} reduces to a joint subcarrier allocation and power allocation problem, which is still an \ac{NP}-hard problem because of the remainder of the subcarrier allocation problem ~\cite{4453890,6514044,8422362}, though with less complexity as compared to the original optimization problem. Consequently, the proposed heuristic-based solution, that is discussed in this section, tackles the original optimization problem in three steps, as in Algorithm~\ref{Algorithm: summary}:
\begin{enumerate}
\item Binding of users to \ac{LEDs}.
\item Determining the user-pairs for each \ac{LED} (i.e.,~user pairing).
\item Optimizing subcarrier(s) allocation to user-pairs in each \ac{LED} and power allocation within each pair (i.e.,~subcarrier allocation and power allocation).
\end{enumerate}

As stated before, the max-min optimization problem in Equations~\eqref{eq:objective-2}--\eqref{eq:constraint-e-2} involves both subcarrier and power allocation; however, since each of the \ac{NOMA} users in each pair will share the same assigned subcarriers with a fixed power per subcarrier, then the subcarrier allocation and the power allocation problems may be done in an iterative manner through both the \ac{SA} algorithm and the bisection method. This is discussed in detail later in this~section.

\begin{algorithm}
\SetAlgoLined
\caption{Overview of the proposed heuristic-based solution.} \label{Algorithm: summary}
{\bf Binding of users to \ac{LEDs}: } Assign users to different \ac{LEDs} in such a way that a user cannot be assigned to more than one \ac{LED};\\
{\bf Determining the user-pairs for each LED:} Apply the \ac{D-NLUPA} algorithm to determine the user-pairs per \ac{LED}. Store the user-pairs in $\boldsymbol{\Gamma}(l_i)$\;
{\bf Optimizing subcarrier(s) allocation to user-pairs in each \ac{LED} and power allocation within each pair:} This task allocates the subcarriers to the user-pairs to maximize the minimum achievable rate of the
served users. We apply the \ac{SA} algorithm with an innovative objective function that allocates the users-pairs to the subcarriers while ensuring an almost uniform data-rates for all the served~users.
\end{algorithm}

\subsection{Binding of Users to LEDs}
\label{subsection: assoc1}

Although a user can receive data using any number of subcarriers, all subcarriers should belong to a single \ac{LED}. Therefore, it is necessary to bind users to \ac{LEDs}. In the VLC literature, there are different approaches to attach users to \ac{LEDs}~\cite{9062301}. In this work, we adopted the strategy to bind the users to the \ac{LEDs} that have the channels of maximum strength with the users. We denote the most suitable \ac{LED} for a user $u_j \in \mathcal{N}$ with $\lambda(u_j)$, and its value can be determined using the following equation:

\begin{align}
\label{equation: assoc1}
\lambda(u_j)= \underset{\forall l_i \in  \mathcal{L}}{\operatorname{argmax}|h_{l_i,u_j},u_j|}.
\end{align}

The imposition is that the \ac{NOMA} scheme, termed in this work
as the \ac{NOMA}-imposed-scheme (i.e.,~two users should share the available subcarrier(s)), requires that the number of users allocated to each \ac{LED} should be even. Otherwise, it is not possible to impose \ac{NOMA} fully, and as a result, some subcarriers are allocated to only one user (this scheme is termed in this work as the ``NOMA-not-imposed scheme''). When the number of users allocated to any \ac{LED} is odd, we can formulate a linear constrained optimization problem and solve it to ensure that each \ac{LED} has an even number of users or zero users. Next, we describe the optimization problem. In Equation \eqref{userseq1}, $\overline{|.|}$ denotes a complement operation, and this operation returns one only when $|\lambda(u_j)-i|=0$. The function $f_1(l_i)$ returns the number of users assigned to the \ac{LED} $l_i$.

\begin{align}
\label{userseq1}
f_1(l_i)= \sum_{j=0}^{N-1}\overline{|\lambda(u_j)-i|}.
\end{align}

The distance of the users from their allocated \ac{LEDs} can be given by:

\begin{align}
\label{usersdist1}
f_2(u_j)= d(u_j,l_{\lambda(u_j)}),
\end{align}
\begin{align}
\label{usersdist2}
f_3(u_j)= \underset{i \in \{0,1,..., L-1\}}{\textnormal{max}}d(u_j,l_i).
\end{align}

In Equations (\ref{usersdist1}) and (\ref{usersdist2}), $f_2(u_j)$ denotes the distance of the user $u_j$ from the \ac{LED} to which it is currently allocated, and $f_3(u_j)$ denotes the maximum distance of $u_j$ from any \ac{LED}. The linear constrained optimization problem is shown below:
\begin{alignat}{3}
&\text{Minimize} &\qquad&\sum_{j=0}^{N-1} \frac{f_2(u_j)}{f_3(u_j)},\label{userdist3}\\
&\text{subject to} &  &  \sum_{i=0}^{L-1}(f_1(l_i)\mod 2) = 0. \label{userdist4}
\end{alignat}

The above linear-constrained optimization problem can be solved using many methods, and in our work, we employed an iterative greedy algorithm~\cite{cormen2009introduction} to solve it. The solution is represented by $\{\lambda(u_0),\lambda(u_1), ..., \lambda(u_{N-1}) \}$, where $\lambda(u_j) \in \{l_0,l_1,...,l_{L-1}\}$. \mbox{Algorithm~\ref{Algorithm: binding}} summarizes the aforementioned approach followed for binding users to \ac{LEDs}. The pseudo-code outlines an iterative method whose termination criterion could be the maximum number of iterations (e.g., 1000 iterations). The iterative method makes a random change in the current binding of users and only lets the new solution replace the current one if the new binding is better than or equal to the current one using \mbox{Equations~\eqref{userdist3} and \eqref{userdist4}}. {The simulation results, shown in Section~\ref{Subsection: Validation and Convergence of the Proposed Heuristic-based Solution}, illustrate that the above-described method is efficient in preventing the LEDs from having an odd number of users with a relatively low number of iterations.}

\begin{algorithm}
\SetAlgoLined
\caption{Algorithm for binding users to \ac{LEDs}.} \label{Algorithm: binding}
\KwIn{$l_i \in \mathcal{L}$, $u_j \in \mathcal{N}$}
\KwOut{$\mathcal{N}_{l_i}$= $\{u_j,$ s.t. $u_j \in \mathcal{N}$, and, $\lambda(u_j)= l_i\}$ (a vector that stores the users of LED $l_i$), \newline
$N_{l_i}$ (the number of users in $\mathcal{N}_{l_i}$)}
Bind the users to the \ac{LEDs} that have the channels of maximum strength with the users based on Equation~\eqref{equation: assoc1}\;
Store the set of users attached to each \ac{LED} in a vector. For example, users attached to LED $l_i$ are stored in vector $\mathcal{N}_{l_i}$, and their sum is stored in $N_{l_i}$\;
\If{\textnormal{NOMA-imposed scheme}=={\textnormal{TRUE}} \&\& $f_1(l_i)$ \textnormal{is odd} $\forall l_i \in  \mathcal{L}$}{
\While{\textnormal{Stopping criterion is not reached}}{
Randomly choose a user currently attached to LED $l_i$ and bind it to a new LED $l_j$ (where, both LED $l_i$ and LED $l_j$ are randomly chosen)\;
Determine the cost of the new solution using Equations (\ref{userdist3}) and (\ref{userdist4})\;
\If{\textnormal{the cost of the new solution is worse than the previous one}}
{
Revert the binding of the selected user to \ac{LED} $l_i$}
}
}
return $\mathcal{N}_{l_i}$, $N_{l_i}$\;
\end{algorithm}

\subsection{Determining of the User-Pairs for Each LED}
\label{subsection: user pairing}

We used the \ac{D-NLUPA} method~\cite{8352621} to determine the user-pairs for each \ac{LED}. The \ac{D-NLUPA} method is efficient in finding user-pairs, and it creates pairs in which the first user has a strong channel (or strongest among users not already part of any pair) with a user of a relatively weaker channel.  Algorithm~\ref{Algorithm: NLUPA1} shows the \ac{D-NLUPA} algorithm. The first step is to sort the users with respect to their channel gains. The second step is a loop that creates pairs of strong and weak users and stores them in $\boldsymbol{\Gamma}(l_i)$. The actions on lines $6$--$7$ only execute if the \ac{NOMA}-not-imposed scheme is deployed and the number of users under any \ac{LED} is odd. We should place the last user (i.e.,~the user having minimum channel strength) into a pair whose weak user is null.

\begin{algorithm} 
\SetAlgoLined
\caption{Algorithm to implement the \ac{D-NLUPA} method of user-pairing.} \label{Algorithm: NLUPA1}
\KwIn{$l_i \in \mathcal{L}$,  $\mathcal{N}_{l_i}$, $N_{l_i}$}
\KwOut{$\boldsymbol{\Gamma}(l_i)$ (a set of all user-pair for the \ac{LED} $l_i$)}
Sort the elements in $\mathcal{N}_{l_i}$ in the descending order based on the strength of their channels with \ac{LED} $l_i$\;
\For{$j$ \textnormal{in} 0 \textnormal{to} $\frac{N_{l_i}}{2}$}{
Create a pair $\mathbb{P}_j$ in which the user at index $j$ is the strong user and the user at index $j+\frac{N_{l_i}}{2}$ is the weak user\;
Insert $\mathbb{P}_j$ in $\boldsymbol{\Gamma}(l_i)$\;
}
\If{\textnormal{NOMA-not-imposed scheme}=={\textnormal{TRUE}} \&\& $N_{l_i}$ \textnormal{is an odd number}}
{Create a pair in which the user at index $N_{l_i}-1$ is the strong user and a null value for the weak user, and insert it into $\boldsymbol{\Gamma}(l_i)$\;}
return $\boldsymbol{\Gamma}(l_i)$\;
\end{algorithm} 
 \unskip

\subsection{Optimizing Subcarrier(s) Allocation to User-Pairs in Each LED and Power Allocation within Each Pair}
\label{Section: Allocating subcarrier(s) to user-pairs}

In this subsection, we discuss the problem of allocating the subcarriers to user-pairs, as well as a solution method. As we mentioned in the previous subsection, for each \ac{LED}, the user-pairs allocated to it are represented by $\boldsymbol{\Gamma}(l_i)$, and each \ac{LED} has up to $K$ subcarriers. The allocation aims to maximize, as well as keep uniformness in the data-rates of the users. We also  propose a novel objective function to achieve the dual goals through applying a single function. In the following, we first discuss our proposed objective function, and then discuss a solution method using the \ac{SA} algorithm. We represent the solution of subcarrier allocation using the following notation:
\begin{align}
\label{solrep}
X= \left[\begin{array}{cccc} x_{0,0}, x_{0,0}, ..., x_{0, K-1}\\
x_{1,0}, x_{1,1}, ...,x_{1,K-1}\\
\ldots, \ldots, \ldots \\
x_{L-1,0}, x_{L-1,1}, ...,x_{L-1,K-1}
\end{array} \right],
\end{align}

\noindent where $x_{i,k}$ is an integer, and $ k \in \{0,..., K-1\}$, $i\in \{0,..., L-1\}$, and $x_{i,k}$ denote the index of the user-pair of $\boldsymbol{\Gamma}(l_i)$ assigned to the $k$-th subcarrier of the \ac{LED} $l_i$. The objective of the subcarrier allocation is to maximize the minimum achievable rate of the served users through finding a decision matrix ($X$), and is given by
\begin{align}
\label{objfun1}
\mathcal{O}(X)= \underset{j\in \{0,1, ..., N-1\} }{\textnormal{min}}R_j, \\ \label{objfunc2}
\text{maximize }\mathcal{O}(X).
\end{align}

We propose two new constraints into the above objective function ($\mathcal{O}(X)$) to help the iterative heuristics to obtain a solution in which the data-rates of all users have the following features:
\begin{enumerate}
\item The data-rates of all users are closer to each other.
\item The users should not have a zero data-rate.
\item The data-rate should be as maximal as possible considering the above two conditions.
\end{enumerate}

To include the above enhancements, we propose the objective function as follows:
\begin{alignat}{3}
&\underset{\mathcal{K}, a_s, a_w}{\text{maximize}} &\qquad&\mathcal{O}(X),\label{objfun3-2}\\
&\text{subject to} &  & \frac{\underset{j \in \{0,1,..., N-1\}}{\textnormal{max}}R_j-\underset{j \in \{0,1, ..., N-1\}}{\textnormal{min}}R_j }{\underset{j \in \{0,1, ..., N-1\}}{\textnormal{max}}R_j} - \mathcal{C}  \leq 0, \label{objcons1}\\
& & & R_j > 0, \ \forall j \in \{0,1, ..., N-1\}. \label{objcons2}
\end{alignat}

{In }the above equations, Equation \eqref{objcons1} ensures that the percentage difference between the maximum and minimum data-rates of any two users should be lesser than a predefined constant $\mathcal{C} \in [0,1]$. Equation \eqref{objcons2} ensures that no user suffers a service outage. 

Finally, we convert the above constrained problem into an unconstrained optimization problem using the penalty method~\cite{10.5555/1841101, 6121687}, as follows:
\begin{gather}
\label{objfun3}
f_{\textnormal{zero}}(R_j)= \begin{cases} 1 \text{, \ if }R_j =0 \\
0 \text{, \ otherwise}
\end{cases}\\
f_{\textnormal{cons}}(R_0, R_1, ...,R_{N-1})= \frac{\sum_{j=0}^{N-1}f_{\textnormal{zero}}(R_j)}{N},\\
f_{\textnormal{max}}(R_0, R_1, ..., R_{N-1})= \underset{j \in \{0,1, ..., N-1\}}{\textnormal{max}}R_j,\\
f_{\textnormal{min}}(R_0, R_1,..., R_{N-1})= \underset{j \in \{0,1, ..., N-1\}}{\textnormal{min}}R_j,\\
f_{\textnormal{diff}}(R_0, R_1, ..., R_{N-1})= \frac{f_{\textnormal{max}}- f_{\textnormal{min}}}{f_{\textnormal{max}}} - \mathcal{C}.
\end{gather}

The new objective function can be given by:
\begin{gather}
\label{newobj}
\mathcal{O'}(X)= \mathcal{O}(X) - \mathcal{P}_1 f_{\textnormal{cons}} - \mathcal{P}_2 f_{\textnormal{diff}}, \\ \label{newobj2}
\text{maximize } \mathcal{O'}(X).
\end{gather}

{ In} the above equation, $\mathcal{P}_1$ and $\mathcal{P}_2$ denote the penalty factors. The function $\mathcal{O}(X)$ denotes a data-rate value of a user-pair, and hence it could be zero or positive, and the rates can vary between different user pairs. Therefore, $\mathcal{P}_1$ is set for the avoidance of solutions with zero minimum data-rates (i.e.,~$f_\textnormal{cons}> 0$), $\mathcal{P}_2$ is set for the avoidance of solutions in which the difference among the data-rates of the users is more than the given value (i.e.,~$f_{\textnormal{diff}}>0$). We prioritize the two constraints by setting $\mathcal{P}_1$ $>>$ $\mathcal{P}_2$. The function ($\mathcal{O'}(X)$) is a \ac{NLIP} problem, which is \ac{NP}-hard in general~\cite{10.5555/2765770}. In this work, we adopted the \ac{SA} algorithm, which is a popular method of solving \ac{NP}-hard optimization problems~\cite{TIAN1995629, doi:10.1002/ett.3850, 8715449}.

Now, we briefly describe the \ac{SA} algorithm for finding an allocation of subcarriers to user-pairs in order to maximize the objective function Equation (\ref{newobj2}). Algorithm~\ref{Algorithm: SAfig1} shows an overview of the \ac{SA} algorithm. The input parameters are as follows: (i)~$T_0$ is the initial temperature of the \ac{SA} algorithm, and its value should be high; (ii)~$\alpha$ is the cooling-rate and its value should be between $(0,1)$, usually, its value is kept very close to $1$, such as $0.97$, $0.99$; (iii)~$M$ indicates the number of iterations in the Metropolis function; (iv)~$\beta$ indicates the increase in the value of $M$. The first four lines in the \ac{SA} algorithm initialize the variable and the current solution ($X_{\textnormal{current}}$). The 'while' loop contains the main algorithm, and the termination criterion could be the maximum time or iterations. The 'while' loop in each iteration calls the Metropolis function, which is responsible for exploring the current solution's neighborhood to improve the current solution ($X_{\textnormal{current}}$).

Algorithm~\ref{Algorithm: SAfig2} shows the Metropolis function. The Neighbor function in line 2 creates a new solution by mutating the current solution. In this work, we employ the random mutation that consists of the following two steps: (i)~Randomly choose a \ac{LED} and a subcarrier of it; and (ii)~change the user-pair already allocated to the selected subcarrier to another user-pair attached to the same \ac{LED}. The new solution ($X_\textnormal{new}$) could replace the current solution based on comparing its objective function values with that of the current solution. The RANDOM in the pseudo-code indicates a random number between $(0,1)$. The readers can refer to~\cite{Sait:1999} for more details on the \ac{SA} algorithm. For the optimal power allocation, the bisection method has been utilized, which can be found in~\cite{7069272}.

\begin{algorithm}
\DontPrintSemicolon
\SetAlgoLined
\caption{Overview of the \ac{SA} algorithm.} \label{Algorithm: SAfig1}
\LinesNumbered
\KwIn{$T_0$, $\alpha$, $\beta$, $M$}
\KwOut{$X$}
Initialize $X$ to a random solution\;
$T= T_0$, $X_{\textnormal{current}}$= $X$, $X_{\textnormal{best}}$= $X$\;
$c_{\textnormal{current}}$ = $\mathcal{O}'(X)$\;
$c_{\textnormal{best}}$= $c_{\textnormal{current}}$\;
\While{\textnormal{stopping criterion is not reached}}{
Call Metropolis($X_{\textnormal{current}}$, $c_{\textnormal{current}}$, $X_{\textnormal{best}}$, $c_{\textnormal{best}}$, $T$, $M$)\;
Replace $M$\ by $\beta \times M$, and $T$ by $\alpha \times T$\;}
return $X_{\textnormal{best}}$\;
\end{algorithm}

\begin{algorithm}
\footnotesize
\DontPrintSemicolon
\SetAlgoLined
\caption{Metropolis function.} \label{Algorithm: SAfig2}
\LinesNumbered
\KwIn{$c_{\textnormal{current}}$, $X_{\textnormal{current}}$, $X_{\textnormal{best}}$, $c_{\textnormal{best}}$, $T$, $M$ }
\KwOut{$X_{\textnormal{current}}$}
\While{$M > 0$}{
$X_{\textnormal{new}}$= Neighbor($X_{\textnormal{current}}$)\;
$c_{\textnormal{new}}$= $\mathcal{O}'(X_{\textnormal{new}})$\;
$\Delta c$= $c_{\textnormal{current}} - c_{\textnormal{new}}$\;
\eIf{$c_{\textnormal{best}} < c_{\textnormal{new}} $}{
$c_{\textnormal{best}}=c_{\textnormal{new}}$, $X_{\textnormal{best}}$= $X_{\textnormal{new}}$\;}
{
\If{${\textnormal{RANDOM}} < e^{-\frac{\Delta c} {T}}$}{
$X_{\textnormal{current}}= X_{\textnormal{new}}$, $c_{\textnormal{current}}$= $c_{\textnormal{new}}$\;}}
Decrease $M$ by 1\;}
return $X_{\textnormal{current}}$\;
\end{algorithm}

\subsection{The Complexity Analysis of the Proposed Heuristic-Based Solution}
\label{Section: Complexity Analysis}
In this section, we briefly discuss the time complexity of the algorithms used to accomplish different tasks of the proposed heuristic-based solution. The binding method, discussed in Section~\ref{subsection: assoc1}, requires the determination of the maximum value in an array which can be implemented using max-heap or linear determination~\cite{cormen2009introduction}. The time complexity of the operation to find the maximum value is $O(L)$, and since the operation should be repeated for each user, therefore, the total complexity of the binding method is $O(NL)$. A brief analysis of the time complexity of the algorithm to implement the \ac{D-NLUPA} method, discussed in Section~\ref{subsection: user pairing}, is as follows. The sorting can be accomplished using heapsort~\cite{cormen2009introduction} that has a complexity of $O(N\textrm{log}N)$, and the creation of user-pairs is a linear operation that has complexity equal to $O(N)$. The time complexity of this step is $O(LN\textrm{log}N)$ because the heap-sort should be performed for each LED. The computation for each \ac{LED} is independent of the others. Therefore, we can perform parallel computations, and in that case, the complexity reduces to $O(N\textrm{log}N)$. We optimized the allocation of user-pairs to the subcarriers using the \ac{SA} algorithm. The complexity of this algorithm is usually expressed in terms of the number of iterations needed for it to converge to its best solution~\cite{WHITLEY1996245}. In the Results and Discussions section, we analyzed the convergence aspects of the \ac{SA} algorithm in detail.

\section{Results and Discussions}
\label{Section:Results and Discussions}

In this section, the obtained results for the achievable max-min user rates in a \ac{C-LiAN} architecture are presented. The user pairing, the sub-carrier allocation, and the power allocation were carried out using the \ac{D-NLUPA}, the proposed \ac{SA} algorithm, and the bisection method, respectively, as detailed before. To assess the performance of the proposed heuristic-based solution, the Monte-Carlo simulation that is averaged over $100$ different users' location realizations is used; specifically, each point in the performance curves is an average of implementations of $100$ different users' location realizations. The proposed heuristic-based solution was implemented in R and C++ using R version $4.0.2$, RStudio version $1.3.1$, and Rcpp. We ran the simulations on a desktop computer with an Intel $2.6$ GHz processor and $128$ GB of memory. In the simulations, an empty room was considered, with a dimension of $5 \times 5 \times 3$, unless otherwise stated. The room was equipped with four \ac{LEDs}, unless otherwise stated, mounted in the ceiling of the room with an equal distance between adjacent LEDs, in a square lattice topology, similar to~\cite{8715449,8254785}. The users were distributed in the room according to the uniform random distribution~\cite{7145863,8574984}.  The \ac{LoS} optical channel model in Equation~\eqref{eq: 1} was considered, where the electrical signal power of each \ac{LED} was set to {$35$ dBm} unless otherwise stated, the default values for half the viewing angle of the \ac{LEDs} and the \ac{FoV} of the \ac{PDs} were set to $60^{\circ}$ and $85^{\circ}$, respectively, unless otherwise stated. The \ac{PDs} faced upward towards the ceiling with an area of $1 \ \textnormal{cm}^2$~\cite{8428395}. The number of subcarriers in each \ac{LED} was set to either $16$ or $32$ subcarriers, and the number of users was set to $20$ users, unless otherwise stated. All network parameters used in the simulation are listed in Table~\ref{Table: Network parameters}. The \ac{SA} algorithm parameters were chosen as, $\beta=1.0005$, $\alpha=0.995$, $M=50$, and $T_0=1.0$. The penalty factors $\mathcal{P}_1$ and $\mathcal{P}_2$, in \eqref{newobj}, were set to $1 \times 10^5$ and $10$, respectively.

\begin{table}[!ht]
\centering
\caption{Network parameters.}
\label{Table: Network parameters}
\begin{tabular}{|m{7.55cm}|m{5.2cm}|}
\hline
Parameter name, notation & Value \\ \hline
The electrical power of the input signal, $P_{e}$ & $[30, 35, 40, 45, 50, 55]$ dBm \cite{8037964,8554281,8428395} \\
Total number of users, $N$ & $[10, 20, 30, 40]$\\ 
Total number of \ac{LEDs}, $L$ & $[4,9]$~\cite{8428395,8254785}\\ 
Total number of subcarriers, $K$ & $[16, 32]$~\cite{8428395, 8715449}\\
Room height & $[3, 5, 7, 9]$ meters\\
Semi-angle at half illumination of the \ac{LEDs}, $\phi_{1/2}$ & $60^{\circ}$~\cite{8715449} \\
\ac{FoV} of the \ac{PDs}, $\Psi_{1/2}$ & $85^{\circ}$~\cite{8428395,6691890} \\
The base-band modulation bandwidth of each \ac{LED}, $B_L$ & $20$ MHz~\cite{8428395}\\
Electrical to optical conversion efficiency, $\iota$ & $3.2$~\cite{8715449}\\
Area of the \ac{PD}, $A_p$ & $1.0$ $\textrm{cm}^2$~\cite{8428395,7037113} \\
Optical to electrical conversion efficiency, $\kappa$ & $0.53$ A/W~\cite{8254785} \\
Equivalent noise power spectral density, $Z_o$ & $1\times 10^{-19} A^2/{\textrm{Hz}}$~\cite{8254785,7360112}\\
Refractive index, $\chi$ & 1.5~\cite{1277847,8254785}\\
Gain of optical filter, $T_s(\psi)$ & $1.0$~\cite{8715449} \\
\hline
\end{tabular}
\vspace{-1.5em}
\end{table}

\subsection{Validation and Convergence of the Proposed {Heuristic-Based Solution}}
\label{Subsection: Validation and Convergence of the Proposed Heuristic-based Solution}

\begin{figure}[!ht]
\centering
\includegraphics[width=0.5\textwidth]{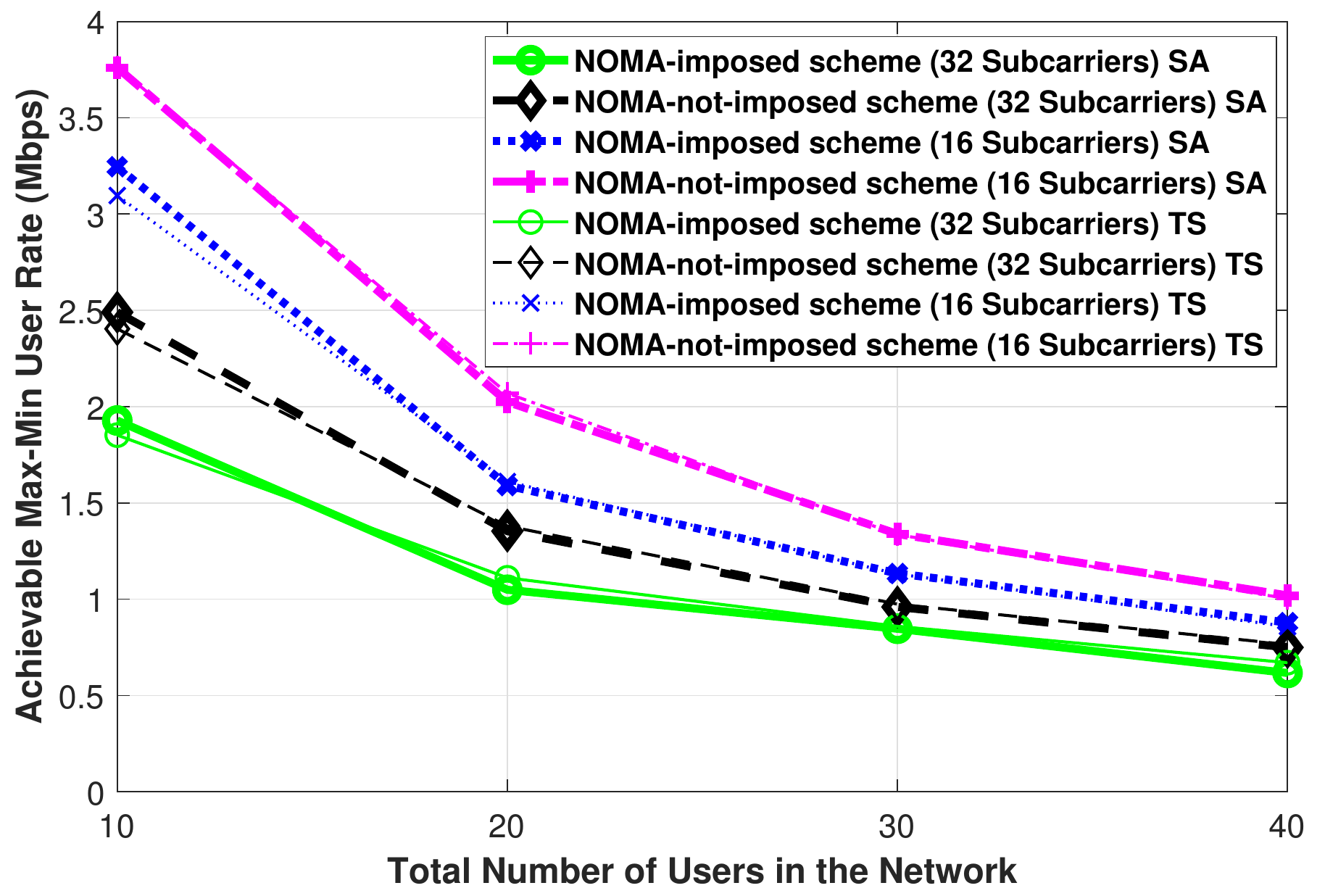}
\caption{The achievable max-min user rate validation curve between the simulated annealing and the Tabu-search algorithms for both the \ac{NOMA}-imposed scheme and the \ac{NOMA}-not-imposed scheme obtained while changing the total number of users in the network. ({$P_e=35$ dBm}, $L=4$, $\phi_{1/2}=60^{\circ}$, $\Psi_{1/2}=85^{\circ}$).}
\label{fig:SA_vs_TS}
\end{figure}

In this subsection, to validate the results that the \ac{SA} algorithm produces for the subcarrier allocation problem, the subcarrier allocation problem is solved using another well-known meta-heuristic, which is the \ac{TS} algorithm that is also efficient in solving non-linear optimization problems~\cite{glover1994tabu}. For validation purposes, both the \ac{SA} and the \ac{TS} algorithms are simulated under the same common parameters. It is worth mentioning that there are two specific parameters to the \ac{TS} algorithm: (i) The ``Tabu-search list'' that includes the most recently visited solutions and set to be $10$, and (ii) the ``Tabu-search-candidate list'' that contains the examined subset of neighborhood solutions and set to be $4$. In addition, we provide the convergence curves of the \ac{SA} algorithm for both the \ac{NOMA}-imposed and the \ac{NOMA}-not-imposed schemes. Fig.~\ref{fig:SA_vs_TS} represents a comparison figure between the Simulated Annealing and the Tabu-search algorithms for both the NOMA-imposed and the NOMA-not-imposed schemes, while changing the total number of users in the network. From this figure, it can be noticed that the obtained results of both algorithms are similar. This can verify that the design, the chosen parameters, and the obtained results for the Simulated Annealing algorithm are suitable and accurate.

\begin{figure}[!ht]
\begin{subfigure}[t]{0.5\textwidth}
\centering
\includegraphics[scale=0.45]{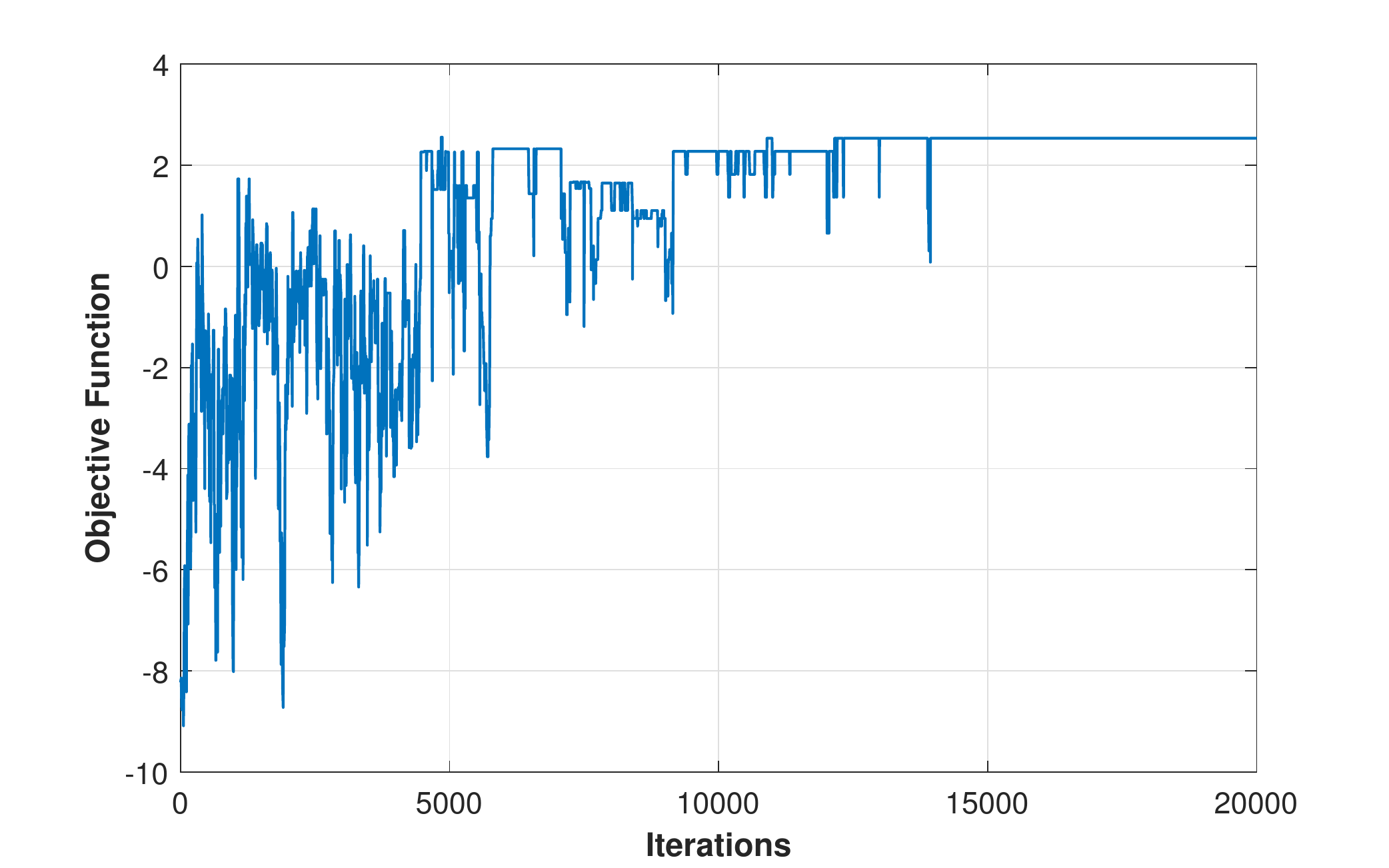}
\caption{NOMA-imposed scheme.}
\label{fig:convergence_Imposed}
\end{subfigure}%
~
\begin{subfigure}[t]{0.5\textwidth}
\centering
\includegraphics[scale=0.45]{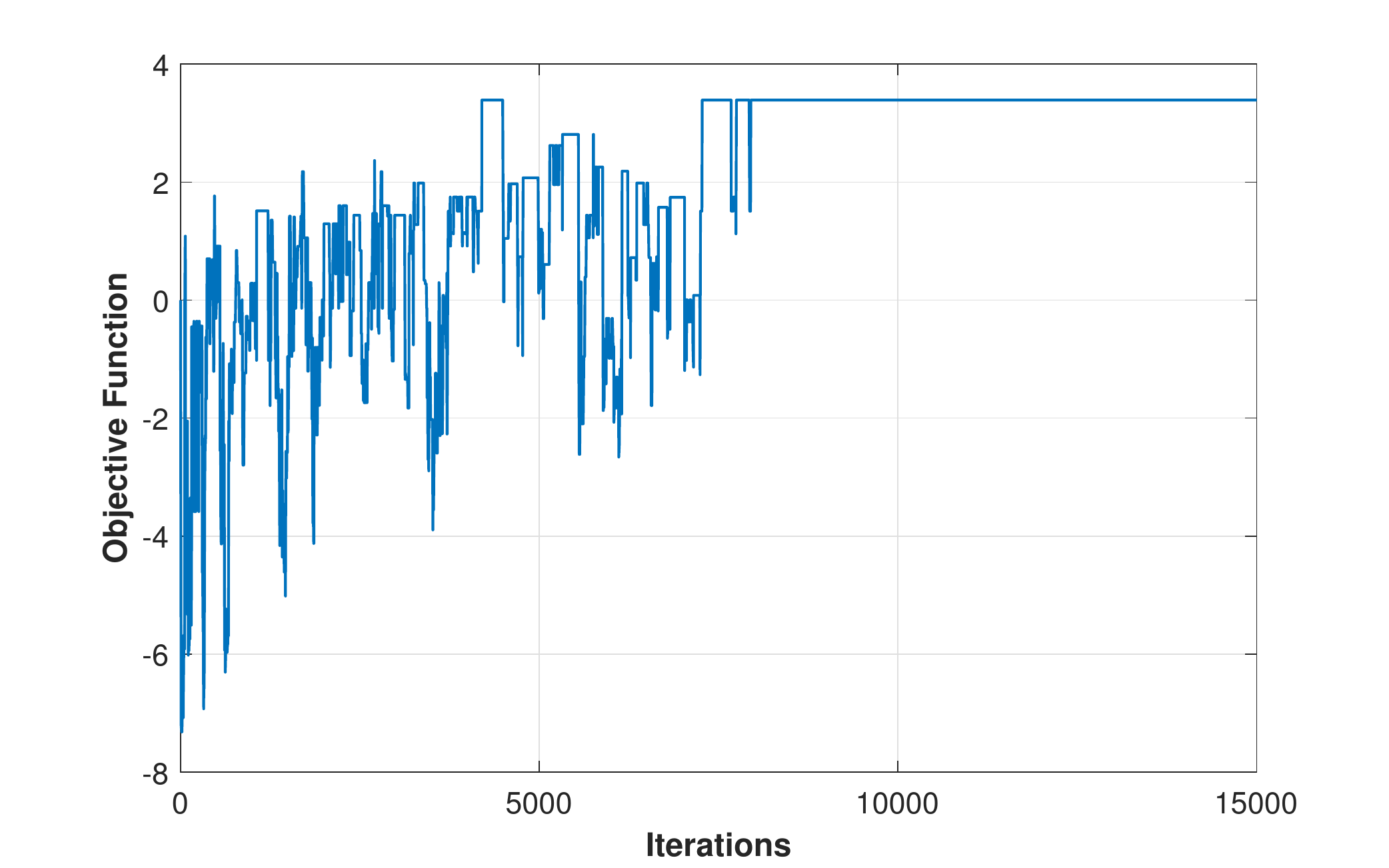}
\caption{NOMA-not-imposed scheme.}
\label{fig:convergence_not_imposed}
\end{subfigure}
\caption{The optimization curves of the \ac{SA} algorithm for both the NOMA-imposed scheme and the NOMA-not-imposed scheme for one users' location realization. ({$P_e=35$ dBm}, $L=4$, $N=10$, $K=16$, $\phi_{1/2}=60^{\circ}$, $\Psi_{1/2}=85^{\circ}$).}
\label{fig:optimcurve}
\end{figure}

Fig.~\ref{fig:optimcurve}a,b shows the optimization convergence curves (i.e.,~objective function versus iterations) of the \ac{SA} algorithm, for the parameter value just mentioned, for both the \ac{NOMA}-imposed scheme and \ac{NOMA}-not-imposed scheme, respectively. The curves indicate that the search process successfully skips through several locally optimal solutions and converges to a good-quality solution. The curve contains negative values due to the violation of constraints. The equations of the objective function Equations~(\ref{newobj}) and (\ref{newobj2}) show that the violation of constraints causes it to return negative values. It should be noted that the number of iterations along the x-axis is equivalent to the number of function evaluations because we computed the objective function once in every iteration. The number of function evaluations is a metric used in \ac{EAs} to denote the convergence and time relationship of \ac{EAs}~\cite{WHITLEY1996245,7892860}. 

\begin{figure}[!ht]
\begin{subfigure}[t]{0.5\textwidth}
\centering
\includegraphics[scale=0.45]{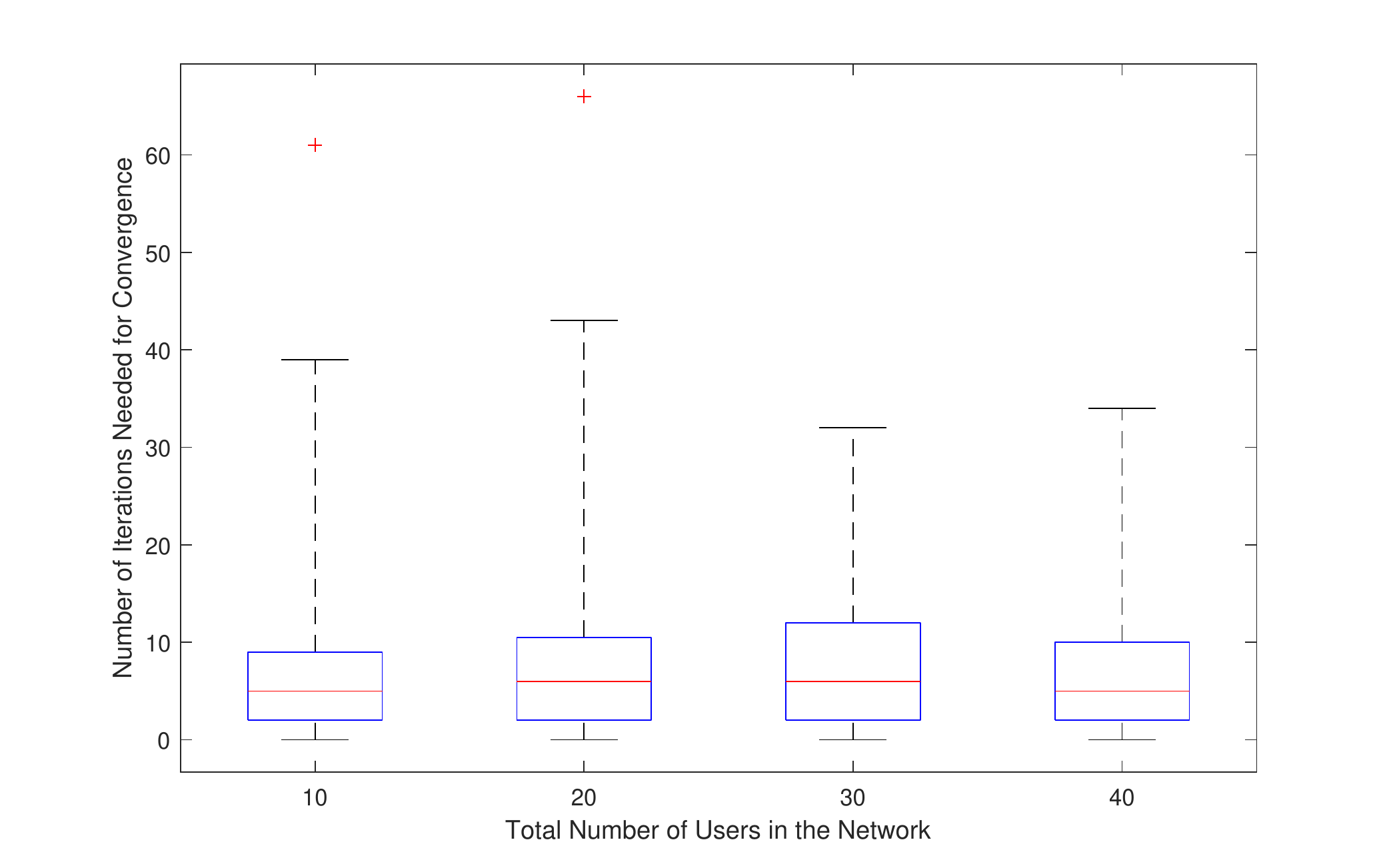}
\caption{}
\label{fig:convergence_boxplot}
\end{subfigure}%
~
\begin{subfigure}[t]{0.5\textwidth}
\centering
\includegraphics[scale=0.45]{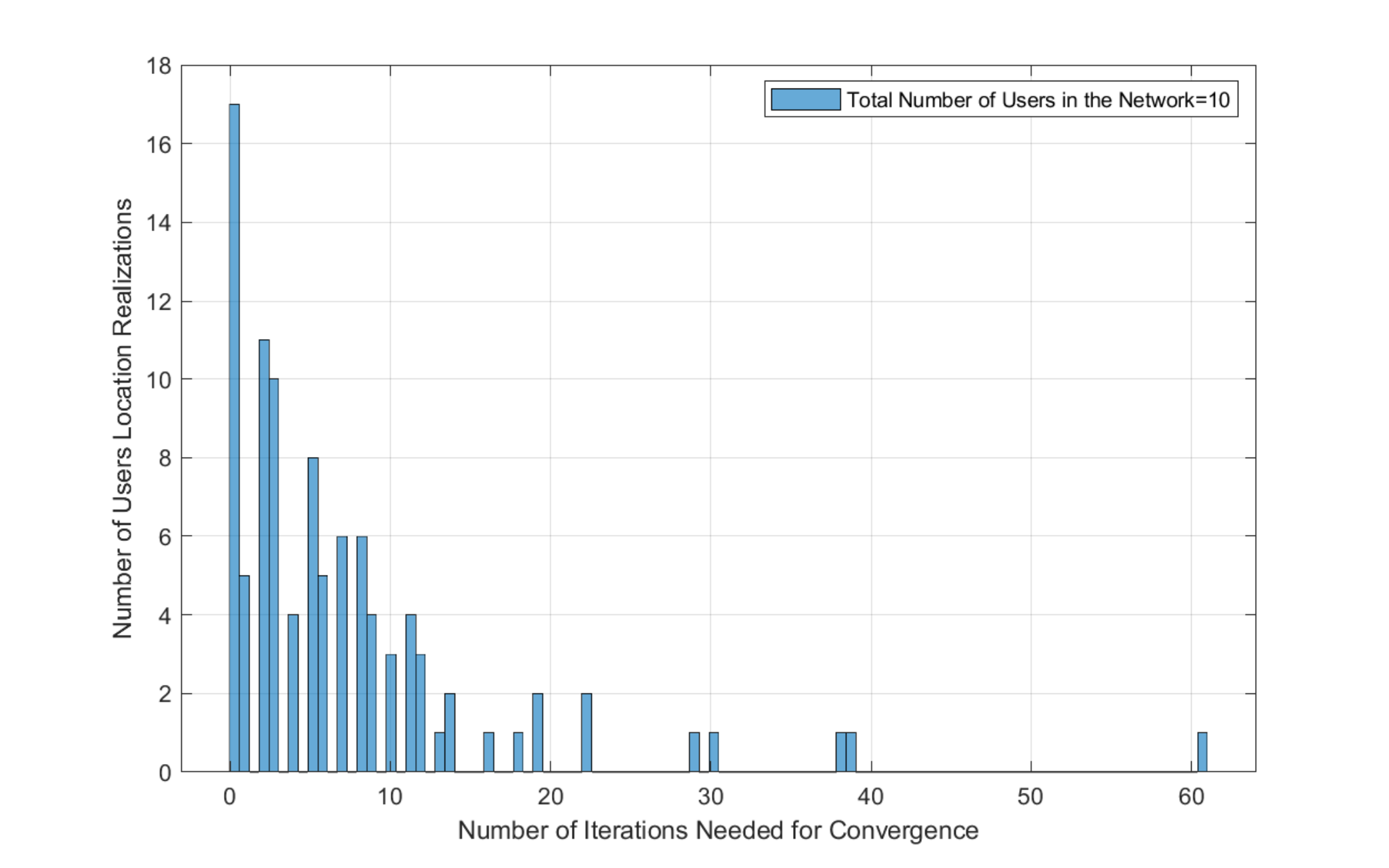}
\caption{}
\label{fig:convergence_hist}
\end{subfigure}
\caption{{ (\textbf{a}) Convergence analysis of the user-binding iterative greedy algorithm. (\textbf{b}) A detailed look into the number of iterations needed for convergence for the test case of $N=10$ in Fig.~\ref{fig:user binding}a. ($P_e=35$ dBm, $L=4$, $K=16$, $\phi_{1/2}=60^{\circ}$, $\Psi_{1/2}=85^{\circ}$, 100 different users' location realizations).}} 
\label{fig:user binding}
\end{figure}

In Fig.~\ref{fig:user binding}, the convergence analysis of the user-binding iterative greedy algorithm is provided. As discussed in Section~\ref{subsection: assoc1}, this algorithm takes place only when the NOMA-imposed scheme is in operation to make sure that the number of users under each LED in the network is even for forming NOMA user-pairs. In Fig.~\ref{fig:user binding}a, box-plot type is used to show the results; therefore, for the reader's convenience, it is necessary to briefly describe the main elements of the box-plot: (i)~The median of the number of iterations needed for convergence (y-axis values) is shown by the red line in the middle of the boxes, (ii)~the ends of the boxes show the lower (Q1) and the upper (Q3) percentile of the y-axis values, (iii)~the small horizontal lines above and beneath the boxes, called the whiskers, show the lowest and the highest value of the number of iterations needed for convergence excluding the outlier points, (iv)~the outlier points are denoted here in a red plus {sign} and refer to the unexpected values. In Fig.~\ref{fig:user binding}a, the number of iterations needed for convergence considering a different number of users in the network, [10,20,30,40] users, is demonstrated. One can observe from the figure that for all considered cases, the algorithm successfully prevents the LEDs from having an odd number of users with a relatively low number of iteration. Comparing the number of iterations needed for convergence in the SA algorithm, provided in Fig.~\ref{fig:optimcurve}, and the number of iterations needed here, we can see that this algorithm converges with a much lower number of iterations. This is related to the reason that the search space for the sub-carrier allocation is much larger than the search space for user-binding. In Fig.~\ref{fig:user binding}b we provide a detailed look into the number of iterations needed for convergence for the test case of $N=10$. In this figure, the convergence is verified using a histogram plot, for 100 users' location realizations, where the number of iterations needed for convergence is shown in the x-axis versus the number of users' location realizations depicted in the y-axis. For example, when the x-axis value equals zero, this indicates that around $17$ {(out of 100)} user realizations have an even number of users under each LED from the beginning (the iterative greedy algorithm is not needed). Next, when the x-axis value equals one, this indicates that {$5$} {(out of 100)} user realizations {need} only one iteration in the iterative greedy algorithm to converge, and so on. This figure consolidates that the greedy algorithm always converges with a small number of iterations for all the considered $100$ uniform random user realizations.

\subsection{The Performance of the Proposed NOMA Schemes}

\begin{figure}[!t]
\begin{subfigure}[t]{0.5\textwidth}
\centering
\includegraphics[scale=0.45]{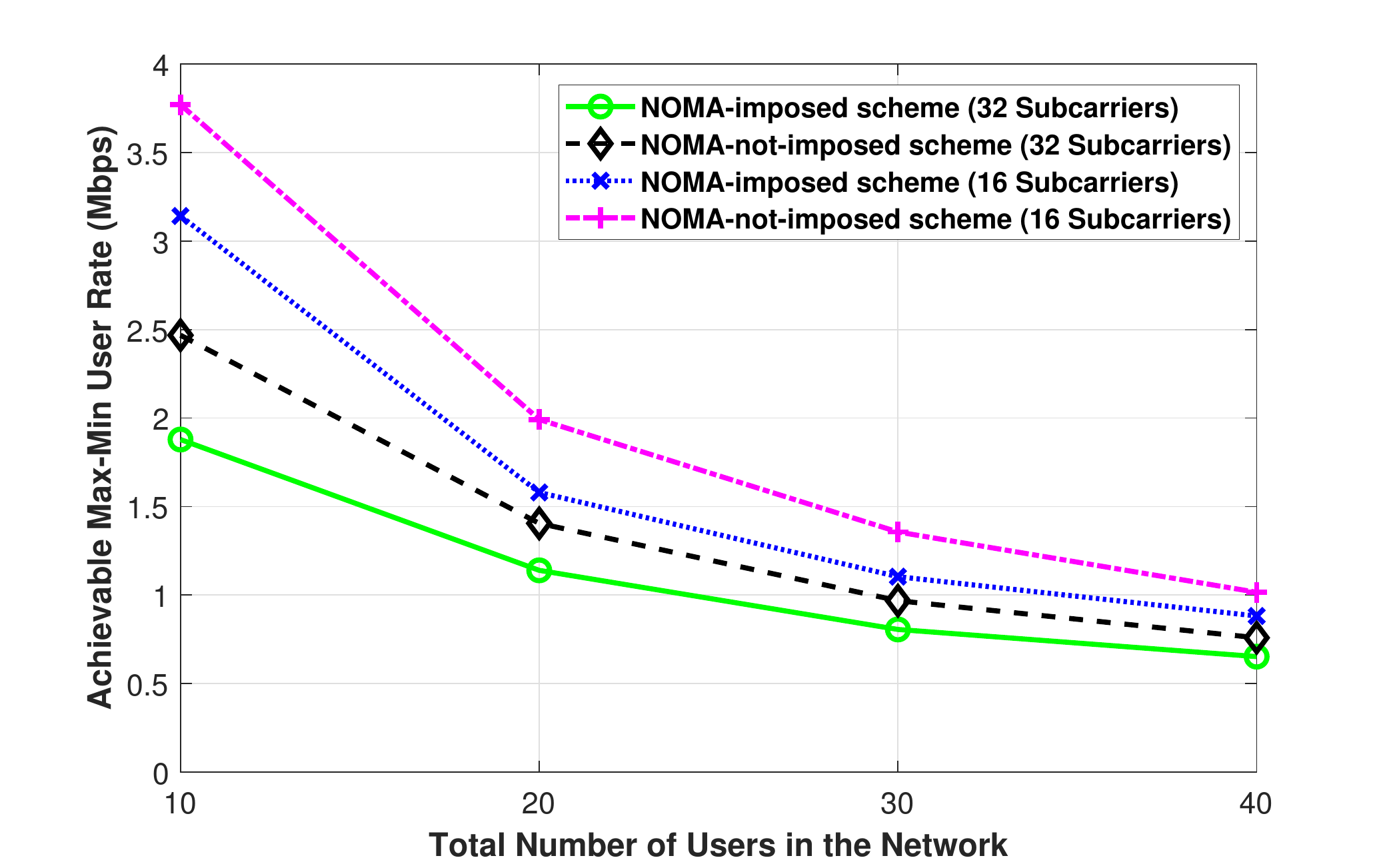}
\caption{For $L=4$.}
\label{fig:2x2}
\end{subfigure}%
~
\begin{subfigure}[t]{0.5\textwidth}
\centering
\includegraphics[scale=0.45]{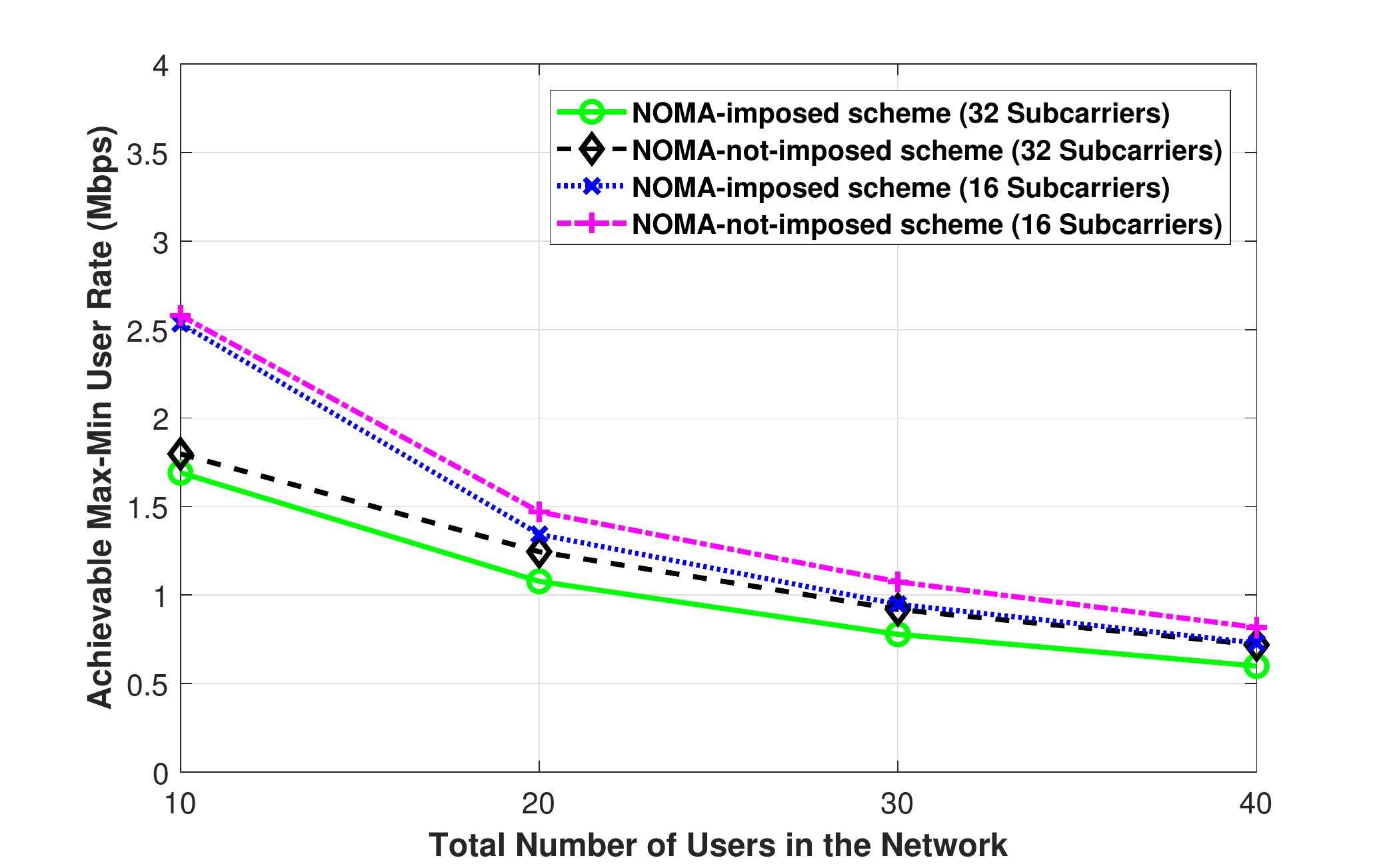}
\caption{For $L=9$.}
\label{fig:3x3}
\end{subfigure}
\caption{The achievable max-min user rate of the \ac{NOMA}-imposed scheme and the \ac{NOMA}-not-imposed scheme obtained while changing the total number of users in the network. ({$P_e=35$ dBm}, $\phi_{1/2}=60^{\circ}$, $\Psi_{1/2}=85^{\circ}$).}
\label{fig:Changing_number_of_users.}
\end{figure}

\begin{figure}
    \centering
    \begin{minipage}{0.485\textwidth}
        \centering
        \includegraphics[scale=0.45]{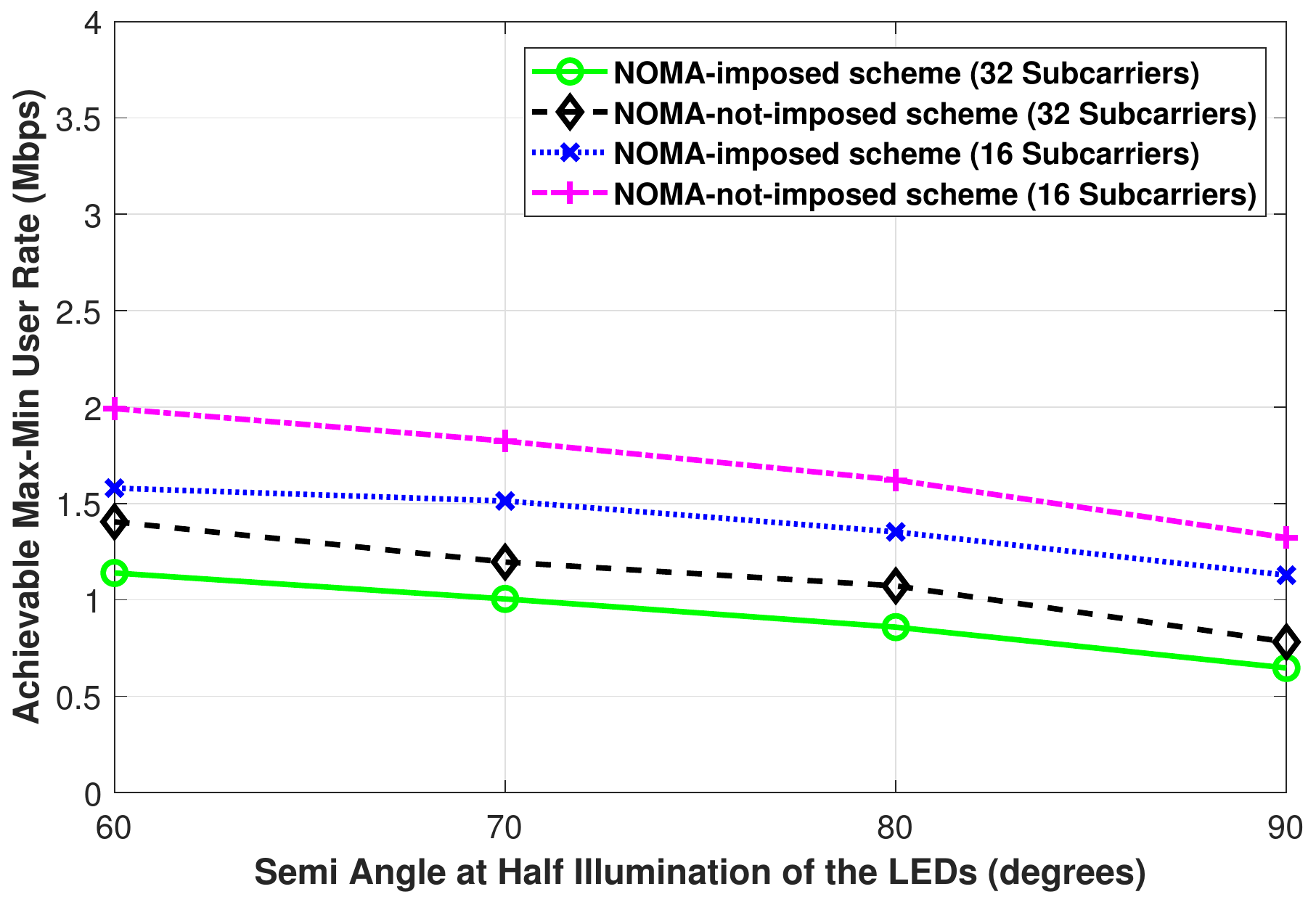} 
        \caption{The achievable max-min user rate of the \ac{NOMA}-imposed scheme and the \ac{NOMA}-not-imposed scheme while changing the semi-angle at half illumination of the \ac{LEDs}. ($P_e=35$ dBm, $L=4$, $N=20$, $\Psi_{1/2}=85^{\circ}$)}
        \label{fig:Changing_LED_semi_angle}
    \end{minipage}\hfill
    \begin{minipage}{0.485\textwidth}
        \centering
        \includegraphics[scale=0.45]{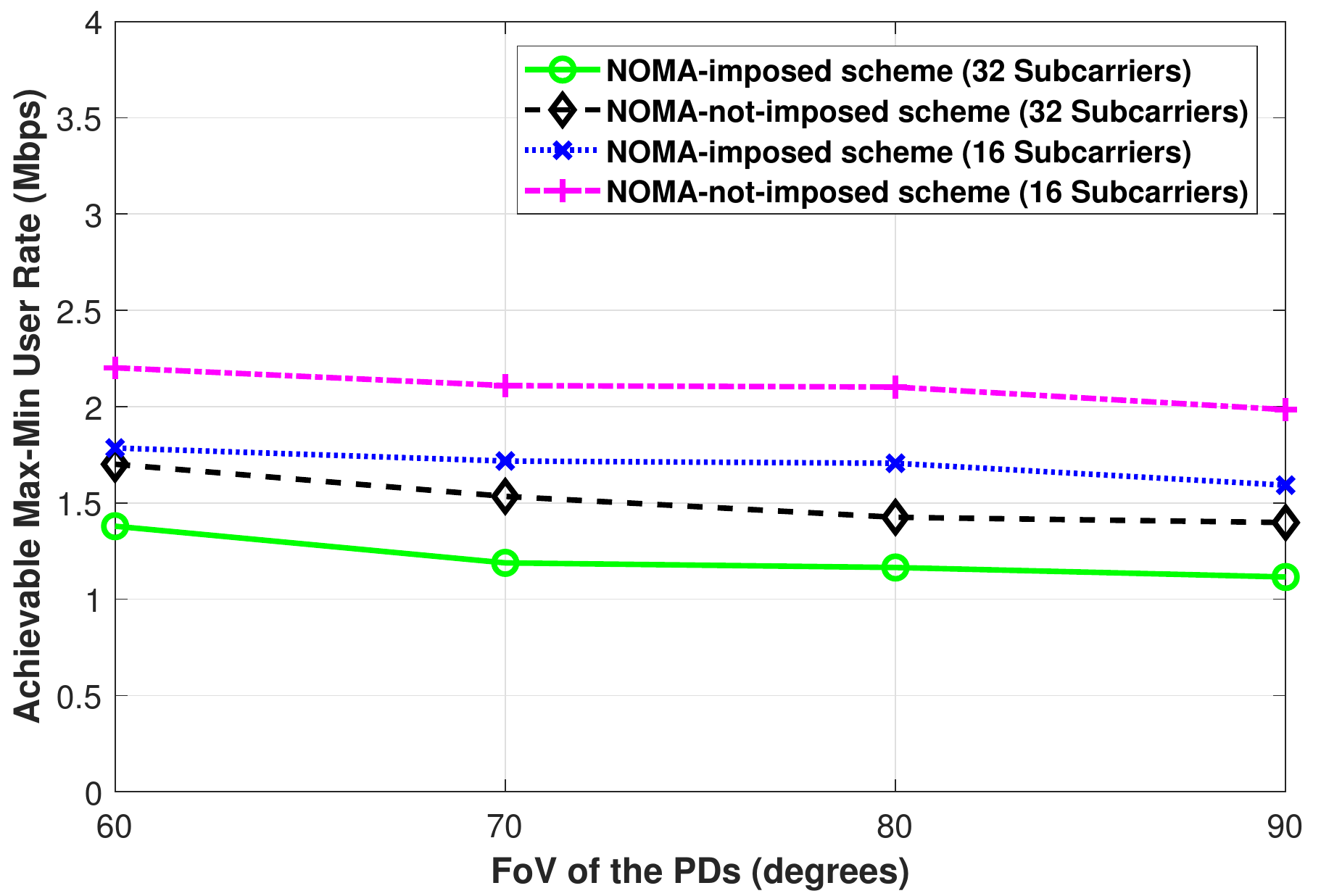} 
        \caption{The achievable max-min user rate of the \ac{NOMA}-imposed scheme and the \ac{NOMA}-not-imposed scheme while changing the \ac{FoV} of the \ac{PDs}. ($P_e=35$ dBm, $L=4$, $N=20$, $\phi_{1/2}=60^{\circ}$)}
        \label{fig:Changing_PD_FoV.}
    \end{minipage}
\end{figure}

\begin{figure}
    \centering
    \begin{minipage}{0.485\textwidth}
        \centering
        \includegraphics[scale=0.45]{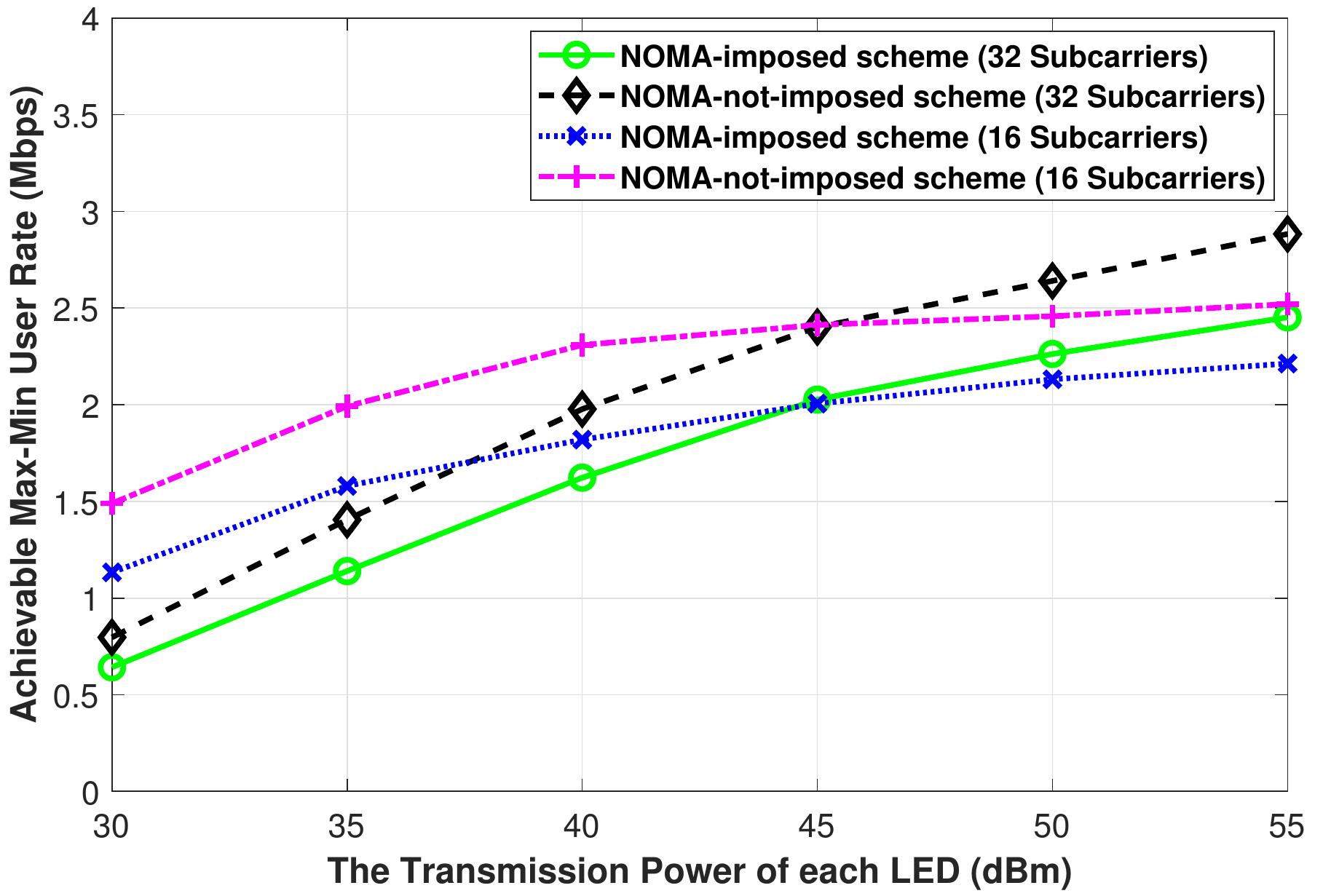} 
        \caption{The achievable max-min user rate of the \ac{NOMA}-imposed scheme and the \ac{NOMA}-not-imposed scheme while changing \ac{LEDs} electrical signal power.  ($L=4$, $N=20$, $\phi_{1/2}=60^{\circ}$, $\Psi_{1/2}=85^{\circ}$)}
        \label{fig:Changing_LED_Tx_power (20users case).}
    \end{minipage}\hfill
    \begin{minipage}{0.485\textwidth}
        \centering
        \includegraphics[scale=0.45]{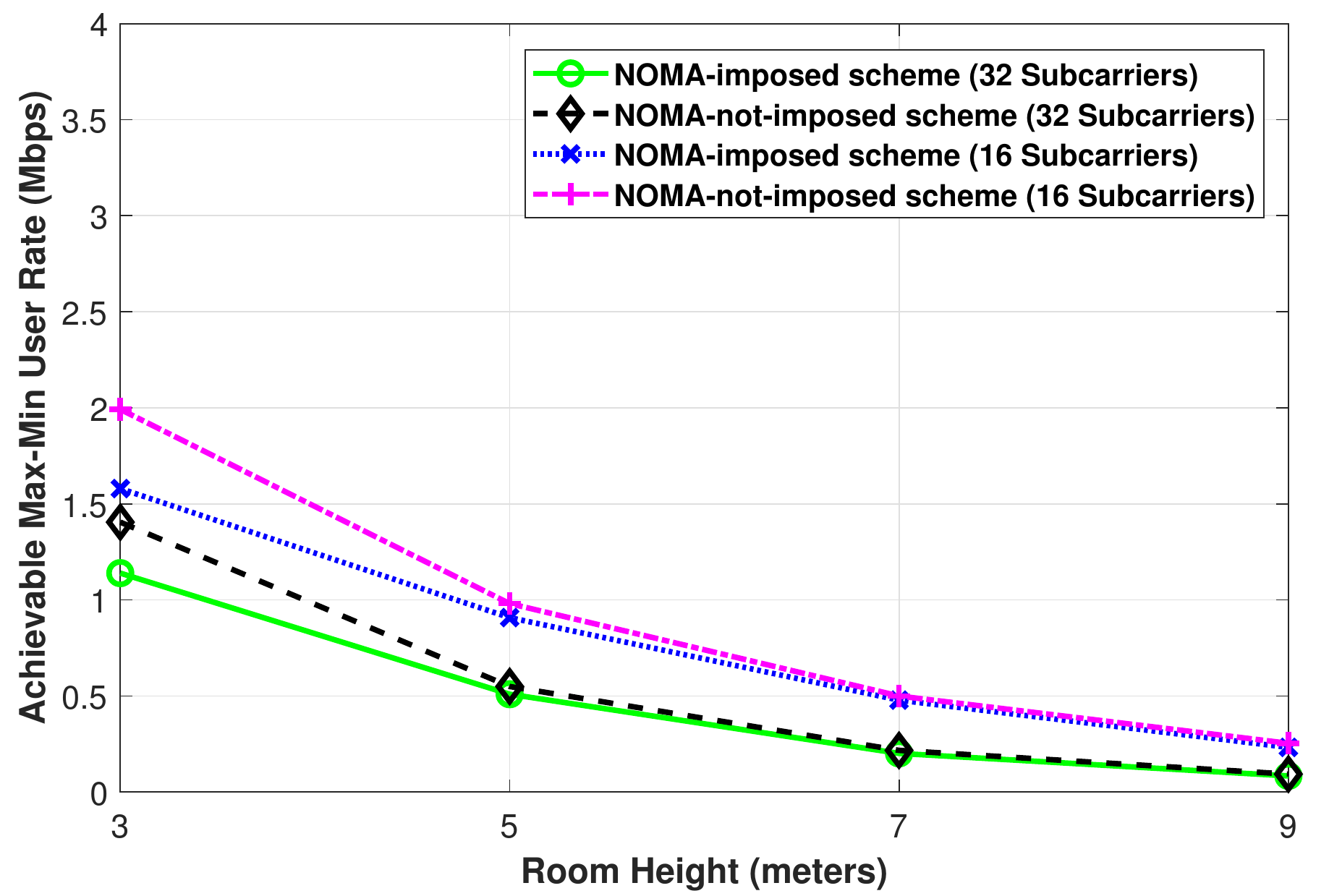} 
        \caption{The achievable max-min user rate of the \ac{NOMA}-imposed scheme and the \ac{NOMA}-not-imposed scheme while changing the room height.  ($P_e=35$ dBm, $L=4$, $N=20$, $\phi_{1/2}=60^{\circ}$, $\Psi_{1/2}=85^{\circ}$) }
        \label{fig:Changing_Room_height.}
    \end{minipage}
\end{figure}

Fig.~\ref{fig:Changing_number_of_users.}a,b illustrates the achievable max-min user rate for the different total number of users, $[10, 20, 30, 40]$ users, and different total number of \ac{LEDs}, $[4,9]$, in the network. The \ac{LEDs} are deployed in a square lattice topology, that is,~2 $\times$ 2 and 3 $\times$ 3. The first observation here, and in the other max-min user rate curves in the other figures (i.e.,~\mbox{Figs~\ref{fig:Changing_LED_semi_angle}--\ref{fig:Changing_Room_height.})}, is that the \ac{NOMA}-imposed scheme tends to achieve lower rates as compared to the \ac{NOMA}-not-imposed scheme. The reason for this is that the \ac{NOMA} scheme's imposition might pair some users with their second-best \ac{LED} in order to impose the \ac{NOMA} scheme fully. The advantage of implementing the \ac{NOMA}-imposed scheme comes from its ability to utilize the resources more efficiently (i.e.,~serving the users with fewer resources) than the \ac{NOMA}-not-imposed scheme that might serve one user in a cluster. The second observation here is that as the number of users in the network increases, the achievable max-min user rate decreases. This is expected, as serving more users by the same resources increases the inter-\ac{LED} interference in the network and results in decreasing the achievable max-min user rates. The third observation here is that the achievable max-min user rates with $32$~subcarriers available per \ac{LED} are less than the achievable max-min user rates with $16$~subcarriers. This is because (i)~the effect of the subcarrier allocation with $16$ subcarriers is more apparent on the achievable max-min user rates compared to the network with $32$~subcarriers; and (ii)~there is a trade-off between the high utilization of subcarriers in the \ac{LEDs} and the interference in the network, in which, if we allocate a high number of subcarriers to users, then the interference in the network will be high, and subsequently, the achievable max-min user rate would decrease. The proposed subcarrier allocation with the \ac{SA} algorithm takes into account this trade-off and eventually, for our setup, the simulations show that the \ac{SINR} and the user rates are better with $16$ subcarriers compared to $32$~subcarriers. Notably, this observation is valid for small to medium values of \ac{LEDs}' electrical signal power \mbox{(i.e.,~{$P_e<45$ dBm})}; however, this trend will change for large $P_e$ values, as discussed later in the discussion of Fig.~\ref{fig:Changing_LED_Tx_power (20users case).}. Finally, one can observe from Fig.~\ref{fig:Changing_number_of_users.}b that the achievable max-min user rate decreases, again, as the number of subcarriers increases, leading to a more interference-limited scenario. Additionally, the performance of both the \ac{NOMA}-imposed and the \ac{NOMA}-not-imposed schemes become closer to each other compared to \mbox{Fig.~\ref{fig:Changing_number_of_users.}a,} which is due to the fact that for a larger number of active \ac{LEDs} in the room, each user can receive strong channels from multiple \ac{LEDs}, as \ac{LEDs} become closer to each other. Subsequently, in the \ac{NOMA}-imposed scheme, the binding of a user to another adjacent \ac{LED} will not change its channel gain, and hence its data rate, much.   

Fig.~\ref{fig:Changing_LED_semi_angle} and Fig.~\ref{fig:Changing_PD_FoV.} show the achievable max-min user rate performance while changing the semi-angle at half illumination of the \ac{LEDs} and changing the \ac{FoV} of the \ac{PDs}, respectively, for both the \ac{NOMA}-imposed scheme and the \ac{NOMA}-not-imposed scheme. In both figures, as the viewing semi-angle of the LED and the \ac{FoV} of the \ac{PDs} increase, a slight decrease in the max-min user rate is observed. This is because, when the viewing semi-angle of the \ac{LED}, as in Fig.~\ref{fig:Changing_LED_semi_angle}, is increased, the coverage region of that \ac{LED} increases, and hence the possibility of binding users that are far from the \ac{LED} increases. On the other hand, in Fig.~\ref{fig:Changing_PD_FoV.}, as the \ac{FoV} of the \ac{PDs} increases, the possibility of binding a user with a \ac{LED} that is far from the user increases. Besides, with a large LED viewing angle, the signal intensity that the user receives from the \ac{LED} is reduced. Likewise, with a large user \ac{FoV}, the user becomes more susceptible to interference from neighbor \ac{LEDs}. A similar trend (i.e., as the \ac{FoV} increases the achievable rate decreases) is reported in~\cite{8554281} for the one-to-many case considering the sum-rate performance metric (in this paper, we consider the max-min user rate metric) in a multi-user \ac{VLC} network.

In Fig.~\ref{fig:Changing_LED_Tx_power (20users case).}, we present the achievable max-min user rate performance while changing the LEDs' electrical signal power (i.e.,~{$P_{e}=[30, 35, 40, 45, 50, 55]$ dBm}) for both the \ac{NOMA}-imposed scheme and the \ac{NOMA}-not-imposed scheme. In this figure, on the $x$-axis, we show the electrical signal power $P_{e}$ but the users' \ac{SINR} and rate equations (i.e.,~\mbox{Equations~\eqref{eq:5} and \eqref{eq:6})} are written in terms of the LED optical transmit power $P_o$. Therefore, we transform the electrical signal power values to optical transmit power values through the following relation $\iota= \frac{P_{o}}{\sqrt{P_{e}}}$ to substitute these values in the simulations. From Fig.~\ref{fig:Changing_LED_Tx_power (20users case).}, one can observe that for small to medium $P_e$ (i.e.,~{$P_e<45$ dBm}) both \ac{NOMA}-imposed and \ac{NOMA}-not-imposed schemes with $16$ subcarriers have a better achievable max-min user rate performance compared to the equivalent schemes with $32$ subcarriers, while the opposite is true for large $P_e$ values (i.e.,~{$P_e \geq 45$ dBm}). The reason is that as the power of the interfering signals gets larger, the achievable max-min user rate tends to have a larger decrease for a smaller number of the allocated subcarriers. A similar trend in \ac{RF} channels, for the secrecy user rate metric, was reported in~\cite{6991587}.

Finally, in Fig.~\ref{fig:Changing_Room_height.}, we illustrate the achievable max-min user rate performance while changing the room height. It is intuitive to see that when the room height increases, the channel gains of the users get weaker and subsequently, the achievable max-min user rate decreases. Besides, with a large room height, the achievable max-min user rates become comparable for both the \ac{NOMA}-imposed scheme or the \ac{NOMA}-not-imposed scheme and with $32$ subcarriers per \ac{LED} or with $16$ subcarriers per \ac{LED}. This is in view of the fact that the users become prone to weak signal power and weak interference power due to the comparable weaker channel gains.

\section{Conclusions}
\label{Section:Conclusions}

As both \ac{VLC} technology and \ac{NOMA}-enabled schemes are envisioned to be among the enablers of a high data rate and low-latency future wireless networks, the achievable max-min user rates of \ac{NOMA}-enabled centralized multi-carrier \ac{VLC} networks were investigated in this paper. The \ac{D-NLUPA} and the bisection method were utilized for user pairing and the power allocation per pair, respectively, and a simulated annealing-based algorithm was developed for the subcarrier allocation. The obtained results, for two variants of the proposed \ac{NOMA}-enabled scheme, have quantified the maximum achievable max-min user rates and shed lights on the effect of the different network parameters, such as: (i) The number of served users, (ii) the number of \ac{LEDs} in the room, (iii) the semi-angle at half illumination of the LEDs, (iv) the FoV of the PDs, (v) the \ac{LED} power, and (vi) the indoor space dimensions on the achievable max-min user rates.

\section{{Extensions and Future Work}}
\label{Section: Extensions and Future Work}

{The} current work in this paper can be extended in the following different directions:
\begin{itemize}
\item By exploiting illuminating LED-arrays, one can enable the utilization of multiple-input multiple-output (MIMO) in indoor VLC networks to extend the network coverage, and further increase the system capacity~\cite{8233180}. Investigating the max-min user rate optimization for indoor MIMO-VLC networks can be considered as a possible direction of future research. However, the performance gains may be limited due to the effect of the peak-to-average power ratio (PAPR) problem~\cite{7506321}.

\item An important practical consideration in indoor VLC networks is user mobility. The Random Way-Point model (RWP) is the most commonly used one for user mobility in indoor VLC literature~\cite{9070153}. In indoor multi-user centralized VLC networks, there are different solutions worth studying which can be adopted to accommodate user mobility: (i)~Horizontal handover while adopting fractional frequency reuse (FFR) scheme or the use of red, green, and blue (RGB) LEDs, or allowing for a coordinated multi-point (CoMP) transmission scheme between different LEDs, (ii)~vertical handover that involves RF/VLC network or WiFi/VLC network or power line communication (PLC)/VLC network, (iii)~cell-zooming strategies that dynamically adjust the coverage areas of the LEDs based on user mobility profiles, and (iv)~utilizing algorithms that can accommodate for user mobility by determining solutions within the coherence time of the channel~\cite{8715449}.
\item A consequence of user mobility in indoor VLC networks is LoS link blockage~\cite{9070153}. Thus, some novel solutions need to be adopted—for example, a multi-directional receiver or omni-directional receiver where PDs are embedded at different sides or all sides, respectively, of a smartphone. Another possible solution for the LoS link blockage can be considered by utilizing intelligent reflecting surfaces (IRSs) inside the indoor environment. Investigating the max-min user rate optimization with such solutions can be an interesting direction of future research.
\end{itemize}

\bibliographystyle{OmarIEEEtran}
\bibliography{Main}

\vskip -2\baselineskip plus -1fil

\begin{IEEEbiography}[{\includegraphics[width=1in,height=1.25in,clip,keepaspectratio]{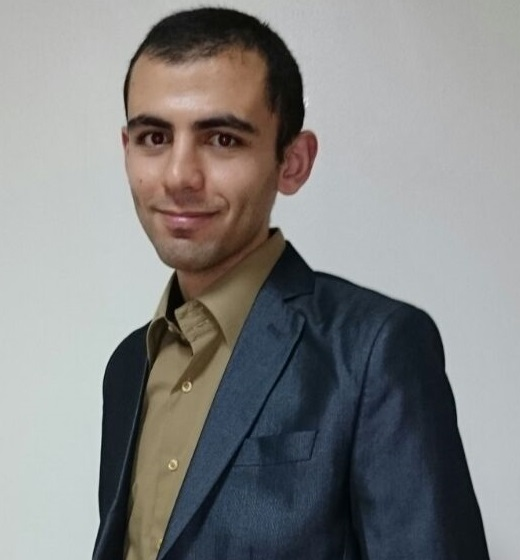}}]{\textbf{Omar Maraqa}} has received his B.S. degree in Electrical Engineering from Palestine Polytechnic University, Palestine, in 2011, and his M.S. degree in Computer Engineering from King Fahd University of Petroleum \& Minerals (KFUPM), Dhahran, Saudi~Arabia, in 2016. He is currently pursuing a Ph.D. degree in Electrical Engineering at KFUPM, Dhahran, Saudi~Arabia. His research interests include performance analysis and optimization of wireless communications systems. \end{IEEEbiography}

\vskip -2\baselineskip plus -1fil

\begin{IEEEbiography}[{\includegraphics[width=1in,height=1.25in,clip,keepaspectratio]{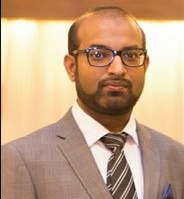}}]{\textbf{Umair F. Siddiqi}} (M'12) was born in Karachi, Pakistan, in 1979. He received the B.E. degree in electrical engineering from the NED University of Engineering and Technology, Karachi, in 2002, the M.Sc. degree in computer engineering from the King Fahd University of Petroleum \& Minerals (KFUPM), Dhahran, Saudi~Arabia, in 2007, and the Dr.Eng. degree from Gunma University, Japan, in 2013. He is currently a Research Engineer with the Center of Communications and Information Technology Research, Research Institute, KFUPM. {He is also currently studying at the University of California San Diego Extension, U.S. in the machine learning certificate program}. He has authored over {40} research papers in international journals and conferences. He holds {five} U.S. patents. His research interests include {machine/deep learning}, metaheuristics, soft computing, and optimization. \end{IEEEbiography}

\vskip -2\baselineskip plus -1fil

\begin{IEEEbiography}[{\includegraphics[width=1in,height=1.25in,clip,keepaspectratio]{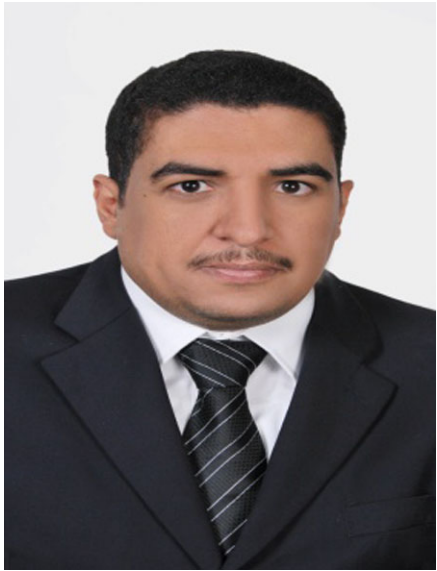}}]{\textbf{Saad Al-Ahmadi}} has received his M.Sc. in Electrical Engineering from King Fahd University of Petroleum \& Minerals (KFUPM), Dhahran, Saudi Arabia, in 2002 and his Ph.D. in Electrical and Computer Engineering from Ottawa-Carleton Institute for ECE (OCIECE), Ottawa, Canada, in 2010. He is currently with the Department of Electrical Engineering at KFUPM as an Associate Professor. His past and current research interests include channel characterization, design, and performance analysis of wireless communications systems and networks.\end{IEEEbiography}

\vskip -2\baselineskip plus -1fil

\begin{IEEEbiography}[{\includegraphics[width=1in,height=1.25in,clip,keepaspectratio]{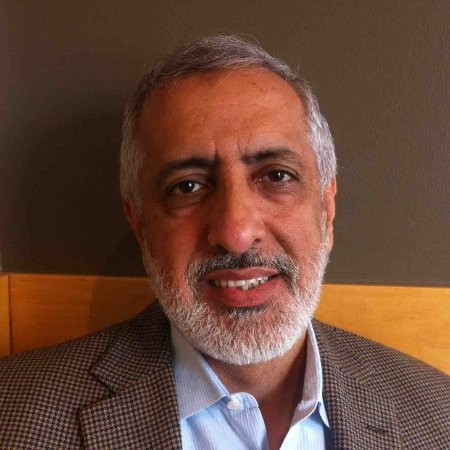}}]{\textbf{Sadiq M. Sait}} (SM'02) was born in Bengaluru. He received the bachelor's degree in electronics engineering from Bangalore University in 1981, and the master's and Ph.D. degrees in electrical engineering from the King Fahd University of Petroleum \& Minerals (KFUPM) in 1983 and 1987, respectively. He is currently a Professor of Computer Engineering and the Director of the Center for Communications and IT Research, Research Institute, KFUPM. He has authored over 200 research papers, contributed chapters to technical books, and lectured in over 25 countries. He is also the Principle Author of two books. He received the Best Electronic Engineer Award from the Indian Institute of Electrical Engineers, Bengaluru, in 1981. \end{IEEEbiography}

\end{document}

%% file: glossaries.tex
\newacronym{SA}{SA}{Simulated Annealing}
\newacronym{VLC}{VLC}{visible light communications}
\newacronym{RF}{RF}{radio frequency}
\newacronym{V2V}{V2V}{vehicle-to-vehicle}
\newacronym{V2X}{V2X}{vehicle-to-everything}
\newacronym{B5G}{B5G}{beyond-fifth generation}
\newacronym{LED}{LED}{light emitting diode}
\newacronym{LEDs}{LEDs}{light emitting diodes}
\newacronym{OMA}{OMA}{orthogonal multiple access}
\newacronym{FDMA}{FDMA}{frequency-division multiple-access}
\newacronym{TDMA}{TDMA}{time-division multiple-access}
\newacronym{CDMA}{CDMA}{code-division multiple-access}
\newacronym{OFDMA}{OFDMA}{orthogonal frequency-division multiple-access}
\newacronym{WDMA}{WDMA}{wavelength-division multiple-access}
\newacronym{NOMA}{NOMA}{non-orthogonal multiple access}
\newacronym{PD-NOMA}{PD-NOMA}{power-domain NOMA}
\newacronym{SC}{SC}{superposition coding}
\newacronym{SIC}{SIC}{successive interference cancellation}
\newacronym{BS}{BS}{base station}
\newacronym{QoS}{QoS}{quality of service}
\newacronym{NP}{NP}{non-deterministic polynomial-time}
\newacronym{DCO-OFDM}{DCO-OFDM}{direct-current biased optical-OFDM}
\newacronym{ITU}{ITU}{international telecommunication union}
\newacronym{FoV}{FoV}{field-of-view}
\newacronym{FoVs}{FoVs}{field-of-views}
\newacronym{CSI}{CSI}{channel state information}
\newacronym{LACO-OFDM}{LACO-OFDM}{layered asymmetrically clipped optical OFDM}
\newacronym{FR}{FR}{frequency reuse}
\newacronym{EAs}{EAs}{evolutionary algorithms}
\newacronym{C-LiAN}{C-LiAN}{centralized light access network}
\newacronym{AP}{AP}{access point}
\newacronym{APs}{APs}{access points}
\newacronym{PD}{PD}{photo-diode}
 \newacronym{PDs}{PDs}{photo-diodes}
\newacronym{SINR}{SINR}{signal-to-noise-interference ratio}
\newacronym{LoS}{LoS}{line-of-sight}
\newacronym{AWGN}{AWGN}{additive white Gaussian noise}
\newacronym{SNR}{SNR}{signal-to-noise ratio}
\newacronym{NLUPA}{NLUPA}{next-largest-difference user-pairing algorithm}
\newacronym{D-NLUPA}{D-NLUPA}{divide-and-next-largest-difference user-pairing algorithm}
\newacronym{NAICS}{NAICS}{network-assisted interference cancellation and suppression}
\newacronym{LTE}{LTE}{long term evolution}
\newacronym{3GPP}{3GPP}{3rd Generation Partnership Project}
\newacronym{CR}{CR}{cognitive radio}
\newacronym{2D}{2D}{two-dimension}
\newacronym{GP}{GP}{gradient projection}
\newacronym{umMTC}{umMTC}{ultra-massive machine-type communication}
\newacronym{IoE}{IoE}{internet-of-everything}
\newacronym{ZF}{ZF}{zero-forcing}
\newacronym{NLIP}{NLIP}{non-linear integer programming}
\newacronym{DP}{DP}{dynamic programming}
\newacronym{MIMO}{MIMO}{multiple-input-multiple-output}
\newacronym{TS}{TS}{Tabu-search}